\pdfoutput=1
\documentclass[12pt,a4paper]{article}

\usepackage{ifthen} 
\newboolean{pdflatex}
\setboolean{pdflatex}{true} 

\newboolean{articletitles}
\setboolean{articletitles}{true} 

\newboolean{uprightparticles}
\setboolean{uprightparticles}{false} 

\def\paperauthors{LHCb collaboration} 
\def\paperasciititle{Template for writing LHCb papers} 
\def\papertitle{Observation and branching fraction measurement of the decay $\Xibm\to\Lb\pim$} 
\def\paperkeywords{{High Energy Physics}, {LHCb}} 
\def\papercopyright{\the\year\ CERN for the benefit of the LHCb collaboration} 
\def\paperlicence{CC BY 4.0 licence}
\def\paperlicenceurl{https://creativecommons.org/licenses/by/4.0/}

\usepackage[top=1in, bottom=1.25in, left=1in, right=1in]{geometry}

\usepackage{microtype}
\usepackage{lineno}  
\usepackage{xspace} 
\usepackage{caption} 

\usepackage{graphicx}  
\usepackage{color}
\usepackage{colortbl}
\graphicspath{{./figs/}} 

\usepackage{amsmath} 
\usepackage{amssymb}
\usepackage{amsfonts}
\usepackage{upgreek} 

\newcommand*\patchAmsMathEnvironmentForLineno[1]{%
\expandafter\let\csname old#1\expandafter\endcsname\csname #1\endcsname
\expandafter\let\csname oldend#1\expandafter\endcsname\csname
end#1\endcsname
 \renewenvironment{#1}%
   {\linenomath\csname old#1\endcsname}%
   {\csname oldend#1\endcsname\endlinenomath}%
}
\newcommand*\patchBothAmsMathEnvironmentsForLineno[1]{%
  \patchAmsMathEnvironmentForLineno{#1}%
  \patchAmsMathEnvironmentForLineno{#1*}%
}
\AtBeginDocument{%
\patchBothAmsMathEnvironmentsForLineno{equation}%
\patchBothAmsMathEnvironmentsForLineno{align}%
\patchBothAmsMathEnvironmentsForLineno{flalign}%
\patchBothAmsMathEnvironmentsForLineno{alignat}%
\patchBothAmsMathEnvironmentsForLineno{gather}%
\patchBothAmsMathEnvironmentsForLineno{multline}%
\patchBothAmsMathEnvironmentsForLineno{eqnarray}%
}

\usepackage{hyperxmp}

\usepackage[pdftex,
            pdfauthor={\paperauthors},
            pdftitle={\paperasciititle},
            pdfkeywords={\paperkeywords},
            pdfcopyright={Copyright (C) \papercopyright},
            pdflicenseurl={\paperlicenceurl}]{hyperref}

\usepackage[colorinlistoftodos,textsize=scriptsize]{todonotes}

\usepackage[bottom,flushmargin,hang,multiple]{footmisc}

\usepackage[all]{hypcap} 

\usepackage{xspace} 
\usepackage{upgreek}

\def\eff{\epsilon}

\def\lhcb   {\mbox{LHCb}\xspace}

\def\MagUp {\mbox{\em Mag\kern -0.05em Up}\xspace}

\ifthenelse{\boolean{uprightparticles}}%
{

 \def\Pmu         {\ensuremath{\upmu}\xspace}

 \def\Ppi         {\ensuremath{\uppi}\xspace}

 \def\Ppsi        {\ensuremath{\uppsi}\xspace}

 \def\PDelta      {\ensuremath{\Delta}\xspace}                 
 \def\PXi         {\ensuremath{\Xi}\xspace}                 
 \def\PLambda     {\ensuremath{\Lambda}\xspace}                 
 \def\PSigma      {\ensuremath{\Sigma}\xspace}                 
 \def\POmega      {\ensuremath{\Omega}\xspace}                 
 \def\PUpsilon    {\ensuremath{\Upsilon}\xspace}
 \let\oldPi\Pi
 \def\PPi         {\ensuremath{\oldPi}\xspace}

 \def\PB      {\ensuremath{\mathrm{B}}\xspace}                 
                  
 \def\PD      {\ensuremath{\mathrm{D}}\xspace}

 \def\PJ      {\ensuremath{\mathrm{J}}\xspace}                 
 \def\PK      {\ensuremath{\mathrm{K}}\xspace}

 \def\Pb      {\ensuremath{\mathrm{b}}\xspace}                 
 \def\Pc      {\ensuremath{\mathrm{c}}\xspace}

 \def\Pi      {\ensuremath{\mathrm{i}}\xspace}

 \def\Ps      {\ensuremath{\mathrm{s}}\xspace}

 \def\thebaroffset{0.0em}
}
{

 \def\Pmu         {\ensuremath{\mu}\xspace}

 \def\Ppi         {\ensuremath{\pi}\xspace}

 \def\Ppsi        {\ensuremath{\psi}\xspace}                 
                  
 \mathchardef\PDelta="7101
 \mathchardef\PXi="7104
 \mathchardef\PLambda="7103
 \mathchardef\PSigma="7106
 \mathchardef\POmega="710A
 \mathchardef\PUpsilon="7107
 \mathchardef\PPi="7105
                  
 \def\PB      {\ensuremath{B}\xspace}                 
                  
 \def\PD      {\ensuremath{D}\xspace}

 \def\PJ      {\ensuremath{J}\xspace}                 
 \def\PK      {\ensuremath{K}\xspace}

 \def\Pb      {\ensuremath{b}\xspace}                 
 \def\Pc      {\ensuremath{c}\xspace}

 \def\Pi      {\ensuremath{i}\xspace}

 \def\Ps      {\ensuremath{s}\xspace}

 \def\thebaroffset{0.18em}
}
\newcommand{\offsetoverline}[2][\thebaroffset]{\kern #1\overline{\kern -#1 #2}}%

\makeatletter
\ifcase \@ptsize \relax
  \newcommand{\miniscule}{\@setfontsize\miniscule{4}{5}}
\or
  \newcommand{\miniscule}{\@setfontsize\miniscule{5}{6}}
\or
  \newcommand{\miniscule}{\@setfontsize\miniscule{5}{6}}
\fi
\makeatother

\DeclareRobustCommand{\optbar}[1]{\shortstack{{\miniscule (\rule[.5ex]{1.25em}{.18mm})}
  \\ [-.7ex] $#1$}}

\def\mup        {{\ensuremath{\Pmu^+}}\xspace}
\def\mun        {{\ensuremath{\Pmu^-}}\xspace} 

\def\squark    {{\ensuremath{\Ps}}\xspace}

\def\cquark    {{\ensuremath{\Pc}}\xspace}

\def\bquark    {{\ensuremath{\Pb}}\xspace}
\def\bquarkbar {{\ensuremath{\overline \bquark}}\xspace}
\def\bbbar     {{\ensuremath{\bquark\bquarkbar}}\xspace}

\def\pion   {{\ensuremath{\Ppi}}\xspace}

\def\pip    {{\ensuremath{\pion^+}}\xspace}
\def\pim    {{\ensuremath{\pion^-}}\xspace}

\def\pims    {{\ensuremath{\pion^-}}\xspace}

\def\kaon    {{\ensuremath{\PK}}\xspace}

\def\KorKbar {\kern \thebaroffset\optbar{\kern -\thebaroffset \PK}{}\xspace}

\def\Kp      {{\ensuremath{\kaon^+}}\xspace}
\def\Km      {{\ensuremath{\kaon^-}}\xspace}

\def\D       {{\ensuremath{\PD}}\xspace}

\def\DorDbar {\kern \thebaroffset\optbar{\kern -\thebaroffset \PD}\xspace}
\def\Dz      {{\ensuremath{\D^0}}\xspace}

\def\Dp      {{\ensuremath{\D^+}}\xspace}
\def\Dm      {{\ensuremath{\D^-}}\xspace}

\def\DpDm    {\ensuremath{\Dp {\kern -0.16em \Dm}}\xspace}

\def\Dstarp  {{\ensuremath{\D^{*+}}}\xspace}

\def\Dsp     {{\ensuremath{\D^+_\squark}}\xspace}

\def\B       {{\ensuremath{\PB}}\xspace}

\def\BorBbar {\kern \thebaroffset\optbar{\kern -\thebaroffset \PB}\xspace}

\def\Bd      {{\ensuremath{\B^0}}\xspace}

\def\BdorBdbar {\kern \thebaroffset\optbar{\kern -\thebaroffset \Bd}\xspace}

\def\Bs      {{\ensuremath{\B^0_\squark}}\xspace}

\def\BsorBsbar {\kern \thebaroffset\optbar{\kern -\thebaroffset \Bs}\xspace}

\def\jpsi     {{\ensuremath{{\PJ\mskip -3mu/\mskip -2mu\Ppsi}}}\xspace}

\def\Y#1S{\ensuremath{\PUpsilon{(#1S)}}\xspace}

\def\Lz          {{\ensuremath{\PLambda}}\xspace}

\def\LorLbar     {\kern \thebaroffset\optbar{\kern -\thebaroffset \PLambda}\xspace}

\def\Sigmares    {{\ensuremath{\PSigma}}\xspace}

\def\Xires       {{\ensuremath{\PXi}}\xspace}
\def\Xiz         {{\ensuremath{\Xires^0}}\xspace}

\def\Lc          {{\ensuremath{\Lz^+_\cquark}}\xspace}

\def\Xicz        {{\ensuremath{\Xires^0_\cquark}}\xspace}

\def\Lb           {{\ensuremath{\Lz^0_\bquark}}\xspace}

\def\Sigmabsm      {{\ensuremath{\Sigmares_\bquark^{(*)-}}}\xspace}
\def\Sigmabspm      {{\ensuremath{\Sigmares_\bquark^{(*)\pm}}}\xspace}

\def\Xib          {{\ensuremath{\Xires_\bquark}}\xspace}

\def\Xibm         {{\ensuremath{\Xires^-_\bquark}}\xspace}

\def\BF         {{\ensuremath{\mathcal{B}}}\xspace}
\def\BR         {\BF}

\def\to                 {\ensuremath{\rightarrow}\xspace}

\def\AT#1     {\ensuremath{A_{\mathrm{T}}^{#1}}\xspace}           

\def\C#1      {\ensuremath{\mathcal{C}_{#1}}\xspace}                       
\def\Cp#1     {\ensuremath{\mathcal{C}_{#1}^{'}}\xspace}                    
\def\Ceff#1   {\ensuremath{\mathcal{C}_{#1}^{\mathrm{(eff)}}}\xspace}        
\def\Cpeff#1  {\ensuremath{\mathcal{C}_{#1}^{'\mathrm{(eff)}}}\xspace}       
\def\Ope#1    {\ensuremath{\mathcal{O}_{#1}}\xspace}                       
\def\Opep#1   {\ensuremath{\mathcal{O}_{#1}^{'}}\xspace}                    

\newcommand{\nospaceunit}[1]{\ensuremath{\text{#1}}}       
\newcommand{\aunit}[1]{\ensuremath{\text{\,#1}}}       

\newcommand{\tev}{\aunit{Te\kern -0.1em V}\xspace}
\newcommand{\gev}{\aunit{Ge\kern -0.1em V}\xspace}
\newcommand{\mev}{\aunit{Me\kern -0.1em V}\xspace}
\newcommand{\kev}{\aunit{ke\kern -0.1em V}\xspace}
\newcommand{\ev}{\aunit{e\kern -0.1em V}\xspace}
 
\newcommand{\mevc}{\ensuremath{\aunit{Me\kern -0.1em V\!/}c}\xspace}
\newcommand{\gevc}{\ensuremath{\aunit{Ge\kern -0.1em V\!/}c}\xspace}
\newcommand{\mevcc}{\ensuremath{\aunit{Me\kern -0.1em V\!/}c^2}\xspace}
\newcommand{\gevcc}{\ensuremath{\aunit{Ge\kern -0.1em V\!/}c^2}\xspace}

\def\mum  {\ensuremath{\,\upmu\nospaceunit{m}}\xspace}

\def\fb   {\ensuremath{\aunit{fb}}\xspace}
\def\invfb   {\ensuremath{\fb^{-1}}\xspace}

\newcommand{\stat}{\aunit{(stat)}\xspace}

\newcommand{\chisq}{\ensuremath{\chi^2}\xspace}

\newcommand{\chisqip}{\ensuremath{\chi^2_{\text{IP}}}\xspace}

\def\gsim{{~\raise.15em\hbox{$>$}\kern-.85em
          \lower.35em\hbox{$\sim$}~}\xspace}
\def\lsim{{~\raise.15em\hbox{$<$}\kern-.85em
          \lower.35em\hbox{$\sim$}~}\xspace}

\def\sPlot{\mbox{\em sPlot}\xspace}

\def\pt         {\ensuremath{p_{\mathrm{T}}}\xspace}

\def\ptot       {\ensuremath{p}\xspace}

\def\evtgen     {\mbox{\textsc{EvtGen}}\xspace}

\def\geant      {\mbox{\textsc{Geant4}}\xspace}

\def\photos     {\mbox{\textsc{Photos}}\xspace}

\def\pythia     {\mbox{\textsc{Pythia}}\xspace}

\def\tell1  {TELL1\xspace}
\def\ukl1   {UKL1\xspace}

\newcommand{\lhcborcid}[1]{\href{https://orcid.org/#1}{\hspace*{0.1em}\raisebox{-0.45ex}{\includegraphics[width=1em]{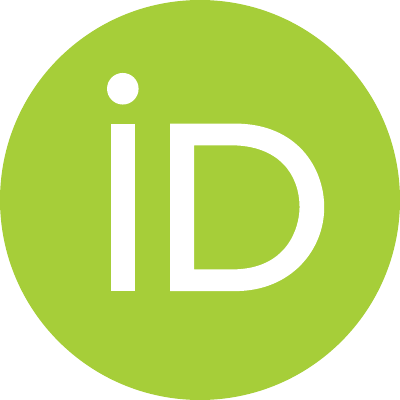}}}}

\usepackage{cite} 
\usepackage{mciteplus}

\usepackage{longtable} 

\begin{document}

\renewcommand{\thefootnote}{\fnsymbol{footnote}}
\setcounter{footnote}{1}


\begin{titlepage}
\pagenumbering{roman}

\vspace*{-1.5cm}
\centerline{\large EUROPEAN ORGANIZATION FOR NUCLEAR RESEARCH (CERN)}
\vspace*{1.5cm}
\noindent
\begin{tabular*}{\linewidth}{lc@{\extracolsep{\fill}}r@{\extracolsep{0pt}}}
\ifthenelse{\boolean{pdflatex}}
{\vspace*{-1.5cm}\mbox{\!\!\!\includegraphics[width=.14\textwidth]{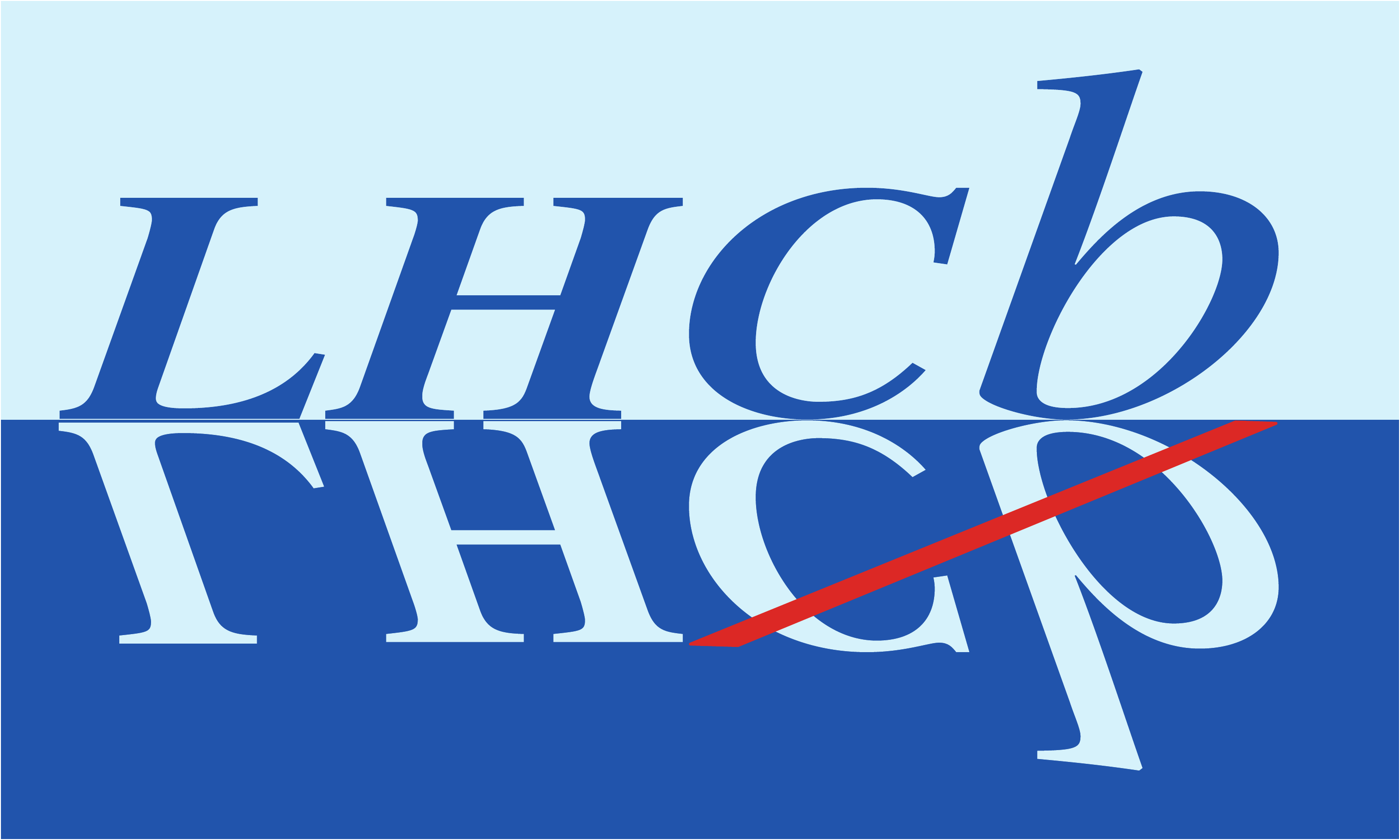}} & &}%
{\vspace*{-1.2cm}\mbox{\!\!\!\includegraphics[width=.12\textwidth]{figs/lhcb-logo.eps}} & &}%
\\
 & & CERN-EP-2023-116 \\  
 & & LHCb-PAPER-2023-015 \\  
 & & Oct 12, 2023 \\ 
 & & \\
\end{tabular*}

\vspace*{2.0cm}

{\normalfont\bfseries\boldmath\huge
\begin{center}
  \papertitle 
\end{center}
}

\vspace*{2.0cm}

\begin{center}
\paperauthors\footnote{Authors are listed at the end of this paper.}
\end{center}

\vspace{\fill}

\begin{abstract}
  \noindent
     The decay $\Xibm\to\Lb\pim$ is observed using a proton-proton collision data sample collected at center-of-mass energy $\sqrt{s}=13$~TeV with the LHCb detector, corresponding to an integrated luminosity of 5.5\invfb. This process is mediated by the $s\to u\bar{u}d$ quark-level transition, where the $b$ quark in the $\Xibm$ baryon is a spectator in the decay.    
    Averaging the results obtained using the two $\Lb$ decay modes, $\Lb\to\Lc\pim$ and $\Lb\to\Lc\pim\pip\pim$, the relative production ratio is measured to be $(f_{\Xibm}/f_{\Lb})\BF(\Xibm\to\Lb\pim)=(7.3\pm0.8\pm0.6)\times10^{-4}$. Here the uncertainties are statistical and systematic, respectively, and $f_{\Xibm}(f_{\Lb})$ is the fragmentation fraction for a $b$ quark into a $\Xibm$ ($\Lb$) baryon. Using an independent measurement of $f_{\Xibm}/f_{\Lb}$, the branching fraction $\BF(\Xibm\to\Lb\pim)=(0.89\pm0.10\pm0.07\pm0.29)\%$ is obtained, where the last uncertainty is due to the assumed SU(3) flavor symmetry in the determination of $f_{\Xibm}/f_{\Lb}$. 
  
\end{abstract}

\vspace*{2.0cm}

\begin{center}
  Published in Phys. Rev. D 108 (2023) 072002
\end{center}

\vspace{\fill}

{\footnotesize 
\centerline{\copyright~\papercopyright. \href{\paperlicenceurl}{\paperlicence}.}}
\vspace*{2mm}

\end{titlepage}


\newpage
\setcounter{page}{2}
\mbox{~}
%
%
%
%


\renewcommand{\thefootnote}{\arabic{footnote}}
\setcounter{footnote}{0}

\cleardoublepage


\pagestyle{plain} 
\setcounter{page}{1}
\pagenumbering{arabic}


\section{Introduction}

In the constituent quark model~\cite{GellMann:1964nj,Zweig:352337} quarks (antiquarks) are $3_C$ ($\bar{3}_{C}$) color-triplets (color-antitriplets) of SU(3)$_{\rm color}$, the gauge group of Quantum Chromodynamics (QCD). Conventional mesons (baryons) are formed from the color-singlet combination of a quark and an antiquark (three quarks).  More complex structures, including tetraquarks, composed of two quarks and two antiquarks, or pentaquarks, formed from four quarks and one antiquark, are expected. These states may be either compact (tightly bound), or molecular (loosely bound). Many candidates of these more exotic states have been reported over the last two decades~\cite{PDG2022,RevModPhys.90.015003}. One natural way to characterize compact tetraquarks and pentaquarks is to describe them as bound states of quarks and diquarks. Diquarks are constructed from two quarks, which together form a color-antitriplet $\bar{3}_C$ of SU(3)$_{\rm color}$. At leading order in QCD, the two quarks in a diquark exhibit an attractive force that is half as strong as that between a quark and an antiquark, when the diquark is in the $J^P=0^+$ state. Thus compact tetraquarks may be formed from the color-singlet combination of a $\bar{3}_C$ diquark and a $3_C$ anti-diquark, and pentaquarks can be built from two $\bar{3}_C$ diquarks and one $\bar{3}_C$ antiquark. In this model, it is possible that some conventional baryons could be best described as the bound state of a $3_C$ quark with a $\bar{3}_C$ diquark. See Refs.~\cite{RevModPhys.65.1199,JAFFE_2005} for reviews on diquarks.

The weak decay $\Xibm\to\Lb\pim$, where the $b$ quark is a spectator, involves the parity-violating $S$-wave matrix element $sd(0^+)\to ud(0^+)+\pim(0^-)$~\cite{GronauPhysRevD.93.034020}. Here, the quantities in parentheses indicate the spin-parity ($J^P$) of the preceding  quark pair or particle. There are several predictions for the size of this matrix element that lead to a branching fraction ${\cal{B}}(\Xibm\to\Lb\pim)$ in the range from 0.14\% to 8\%~\cite{GronauPhysRevD.93.034020,PenyYuNiuPLB826,Cheng2016,FALLER2015653,PhysRevD.90.033016,Voloshin:2000et,PhysRevD.105.094011,PhysRevD.106.093005}. The largest of these predictions~\cite{PhysRevD.90.033016,Voloshin:2000et} is derived from a current algebra approach, and considers the possibility of enhanced short-range correlations within diquarks~\cite{Shifman:2005wa, Dosch:1988hu} that could significantly increase the hadronic matrix element for the $\Xibm\to\Lb\pim$ transition, leading to a value of ${\cal{B}}(\Xibm\to\Lb\pim)$ in the range of (2--8)\%. If such a large branching fraction was measured, it would be among the largest of any $\Xibm$ decay mode.  

In the absence of any large enhancements to the $\Xibm$ decay rate from the $sd$ diquark, a rough estimate of the ratio of the $s$ quark decay rate to the $b$ quark decay rate within the $\Xib$ baryon can be obtained from the ratio of lifetimes $\tau_{\Lb}/\tau_{\Lz}\simeq0.58\%$~\cite{PDG2022}. Thus, a branching fraction for the decay $\Xibm\to\Lb\pim$ anywhere in the range of $(2-8)$\% would be rather striking, and could lend support to diquark models and enhanced short-range correlations within diquarks.

Regardless of the value for the $\Xibm\to\Lb\pim$ decay rate, this contribution should be accounted for when comparing the measured $\Xibm$ lifetime to theoretical predictions, such as those made using the Heavy Quark Expansion (see Ref.~\cite{Lenz:2014jha} for a review). Such comparisons typically consider only the decay of the heavy $b$ quark, along with higher-order corrections due to the spectator quarks, and do not include contributions from the decay of the $s$ quark. 

A previous search for the $\Xibm\to\Lb\pim$ decay was performed by the LHCb collaboration using 3\invfb of proton-proton ($pp$) collision data at center-of-mass energies $\sqrt{s}=7$ and 8 TeV, and found evidence for the decay at the level of 3.2\,$\sigma$ significance~\cite{LHCb-PAPER-2015-047}. This paper reports a follow-up analysis of this decay mode, and measurement of the branching fraction $\BF(\Xibm\to\Lb\pim)$. The inclusion of charge-conjugate processes is implied throughout. The measurement of $\BF(\Xibm\to\Lb\pim)$ is performed by normalizing the $\Xibm$ signal yield to the yield of inclusively produced $\Lb$ baryons through the equation 
\begin{align}
r_s\equiv\frac{f_{\Xibm}}{f_{\Lb}}\BR(\Xibm\to\Lb\pim) = \frac{N(\Xibm\to\Lb\pim)}{N(\Lb)~\eff_{\rm rel}}.
\end{align}
\noindent Here $f_{\Xibm}$ and $f_{\Lb}$ are the $b\to\Xibm$ and $b\to\Lb$ fragmentation fractions, respectively, with the ratio measured by LHCb to be $\frac{f_{\Xibm}}{f_{\Lb}}=(8.2\pm0.7\pm0.6\pm2.5)$\%~\cite{LHCb-PAPER-2018-047} at $\sqrt{s}=13$~TeV. The yield in the normalization mode, $N(\Lb)$, is the number of selected $\Lb$ signal decays from all sources, and
$N(\Xibm\to\Lb\pim)$ is the yield of $\Xibm\to\Lb\pim$ signal decays. The factor $\eff_{\rm rel}\equiv\eff_{\rm sig}/\eff_{\rm norm}$ is the relative efficiency between the signal and normalization modes.  

The measurement uses $pp$ collision data samples collected by the LHCb experiment at $\sqrt{s}=13$~TeV, corresponding to
an integrated luminosity of 5.5\invfb. The integrated luminosity and $\bbbar$ production cross-section are each about a factor of two larger compared to the previous analysis~\cite{LHCb-PAPER-2015-047}. Moreover, the previous work used the single decay mode, $\Lb\to\Lc\pim$, and this analysis uses both $\Lb\to\Lc\pim$ and $\Lb\to\Lc\pim\pip\pim$ decays. Taken together, a substantial improvement in statistical precision is expected compared to the previous measurement.

The signal for the $\Xibm\to\Lb\pim$ decay is a narrow peak in
the spectrum of the mass difference, $\delta M\equiv M(\Lb\pim)-M(\Lb)-m_{\pi}$, where $M(\Lb\pim)$ and $M(\Lb)$ are the reconstructed invariant masses of the respective candidates, and $m_{\pi}$ is the known $\pim$ mass~\cite{PDG2022}. From the known masses of the $\Xibm$~\cite{LHCb-PAPER-2020-032} and $\Lb$ baryons~\cite{PDG2022}, the peak is expected at $38.14\pm0.29$\mevcc.

\section{Detector and simulation}

The \lhcb detector~\cite{LHCb-DP-2008-001,LHCb-DP-2014-002} is a single-arm forward
spectrometer covering the forward direction and
designed for the study of particles containing \bquark or \cquark
quarks. The detector includes a high-precision tracking system
consisting of a silicon-strip vertex detector surrounding the $pp$
interaction region~\cite{LHCb-DP-2014-001}, a large-area silicon-strip detector located
upstream of a dipole magnet with a bending power of about
$4{\mathrm{\,Tm}}$, and three stations of silicon-strip detectors and straw
drift tubes~\cite{LHCb-DP-2017-001} placed downstream of the magnet.
The tracking system provides a measurement of the momentum, \ptot, of charged particles with
a relative uncertainty that varies from 0.5\% at low momentum to 1.0\% at 200\gevc.
The minimum distance of a track to a primary $pp$ collision vertex (PV), the impact parameter (IP), 
is measured with a resolution of $\sigma_{\rm IP}=(15+29/\pt)\mum$,
where \pt is the component of the momentum transverse to the beam, in\,\gevc.
Different types of charged hadrons are distinguished using information
from two ring-imaging Cherenkov detectors~\cite{LHCb-DP-2012-003}. 
Photons, electrons and hadrons are identified by a calorimeter system consisting of
scintillating-pad and preshower detectors, an electromagnetic
and a hadronic calorimeter. Muons are identified by a
system composed of alternating layers of iron and multiwire
proportional chambers~\cite{LHCb-DP-2012-002}.
The online event selection is performed by a trigger~\cite{LHCb-DP-2019-001}, 
which consists of a hardware stage, based on information from the calorimeter and muon
systems, followed by a software stage, which applies a full event
reconstruction. The software stage employs a multivariate algorithm~\cite{BBDT,LHCb-PROC-2015-018} to
identify secondary vertices consistent with the decay
of a \bquark hadron.

Simulation is required to model the effects of the detector acceptance and the
imposed selection requirements. In the simulation, $pp$ collisions are generated using
\pythia~\cite{Sjostrand:2007gs,*Sjostrand:2006za} 
with a specific \lhcb configuration~\cite{LHCb-PROC-2010-056}.
Decays of unstable particles are described by \evtgen~\cite{Lange:2001uf}, in which final-state radiation is generated using \photos~\cite{davidson2015photos}.
The interaction of the generated particles with the detector, and its response, are implemented using the \geant toolkit~\cite{Allison:2006ve, *Agostinelli:2002hh} as described in Ref.~\cite{LHCb-PROC-2011-006}. The underlying $pp$ interaction is reused multiple times, with an independently generated signal decay for each one~\cite{LHCb-DP-2018-004}.

\section{Event selection}

Two samples of $\Lb$ candidates are reconstructed, one composed of $\Lc\to p\Km\pip$ candidates combined with a single $\pim$ (1$\pi$ sample), and a second formed by pairing $\Lc$ candidates with three pions, $\pim\pip\pim$~($3\pi$ sample). All final-state tracks are required to be significantly detached from all PVs in the event, pass loose particle identification (PID) criteria for the particle to be consistent with the assumed decay, and have $\pt>100$\mevc. The $\Lc$ baryon is required to have $|M(p\Km\pip)-m_{\Lc}|<25$\mevcc, where $m_{\Lc}$ is the known $\Lc$ mass~\cite{PDG2022}, and have a decay time in the range $-0.5<t<5$~ps. For the latter, the negative lower bound allows for finite resolution of the $\Lc$ baryon decay time.
For the $\Lb\to\Lc\pim\pip\pim$ decay, the dominant resonant feature of the $\pim\pip\pim$ system is the contribution from the broad $a_1(1260)^-$ resonance. Since the combinatorial background under the $\Lb$ baryon signal peak rises steeply with $M(\pim\pip\pim)$, a requirement of $M(\pim\pip\pim)<2800$\mevcc is imposed, which retains about 90\% of the signal. Each $\Lb$ candidate is assigned to the PV  for which $\chisqip$ is smallest, where $\chisqip$ is the increase in $\chisq$ of the PV fit when the particle under consideration is included in the fit. To a good approximation, $\chisqip\simeq({\rm IP}/\sigma_{\rm IP})^2$, and all decay products of the $\Lb$ candidate are required to have $\chisqip>4$.

Backgrounds from other $b$-hadron decays, where the $\Lc$ candidate is compatible with a misidentified $D$ meson, are suppressed by re-computing the three-body mass with the $\Dp\to\Km\pip\pip,\Kp\Km\pip$, $\Dsp\to\Kp\Km\pip$, $\Dstarp\to(\Dz\to\Kp\Km)\pip$ mass hypotheses, and applying stringent PID requirements if the mass is consistent with any of the above charm mesons within about twice its resolution. A similar procedure is employed to remove the contribution from $\phi\to\Kp\Km$ decays where the $\Kp$ is misidentified as a proton. The efficiency of this veto is about 98\% on simulated signal decays, while removing about 15\% of the misidentified  backgrounds.

In a small fraction of cases, a single particle in the vertex detector can be mis-reconstructed as two distinct tracks that have almost zero opening angle. 
This background is removed by requiring all pairs of final-state tracks in the decay to have an opening angle larger than 0.8 mrad. The efficiency of this requirement on simulated $\Lb$ decays is 99.8\%, while suppressing the mis-reconstruction background by 3\% (14\%) for the $\Lb\to\Lc\pim$ (${\Lb\to\Lc\pim\pip\pim}$) candidates. 

The $\Xibm$ sample is formed by combining each $\Lb$ candidate satisfying ${5560<M(\Lb)<5680~\mevcc}$ with a $\pims$ meson candidate. The $\pim$ meson must pass loose PID requirements, have $\pt>100\mevc$, and have an opening angle larger than 0.8~mrad relative to each of the other final-state particles of the $\Xibm$ candidate. No requirement is imposed on $\chisqip$ for the $\pim$ meson, because its
low $\pt$, about 0.35\gevc on average for those pions that are reconstructed, leads to a relatively large uncertainty on $\sigma_{\rm IP}$. The fitted $\Xibm$ decay vertex is required to have good fit quality. Using the same selection criteria as the right-sign (RS) $\Lb\pim$ combinations, wrong-sign (WS) combinations are also formed to study the combinatorial background, and to train the multivariate discriminants discussed below. When there are multiple candidates in an event, which happens in a few percent of selected events, all candidates are retained.
The $\Lb$ and $\Xibm$ baryons are required to be within the fiducial region $\pt<20$\gevc and $2<\eta<6$, which is the region in which $f_{\Xibm}/f_{\Lb}$ was measured~\cite{LHCb-PAPER-2018-047}.

To improve the signal-to-background ratio in the $\Lb$ normalization and $\Xibm$ signal modes, two pairs of boosted decision tree (BDT) classifiers~\cite{Breiman, AdaBoost, TMVA4} with gradient boosting are employed. The first pair (BDT1), one applied to the $1\pi$ and the second for the 3$\pi$ mode,
is used to suppress the combinatorial background under the $\Lb$ signal peak. 
The second pair (BDT2) suppresses the combinatorial background under the $\Xibm$ signal peak. Each classifier has an output variable that ranges from $-1$ to 1, with the signal (background) events peaking toward 1 ($-1$).

The optimization of these BDT algorithms requires accurate determination of the efficiency from simulation. Two weights are applied to the simulated events to account for the differences in the kinematical distributions in the final state between data and simulation. The first weight accounts for the differences in the $(\pt,\eta)$ production spectra of the beauty baryons. For the $\Lb$ decays, the weights are obtained from the ratio of the background-subtracted $(\pt,\eta)$ distributions in $\Lb\to\Lc\pim(\pip\pim)$ data using the \sPlot method~\cite{Pivk:2004ty} and simulated signal decays.
For the $\Xibm$ simulation, $(\pt,\eta)$ weights are obtained from the $\Xibm\to\Xicz\pim$ decay, where $\Xicz\to p\Km\Km\pip$. 

The second weight accounts for imperfect modeling of the resonant contributions to the $\Lc\to p\Km\pip$ decay. This weight is a function of the pair of invariant masses $[m(p\Km), m(\Km\pip)]$ in the $\Lc$ decay, and are obtained from large samples of semileptonic $\Lb\to\Lc\mu\nu_{\mu}X$ decays in data and simulation. Both weights are applied when computing all efficiencies.

The BDT1 classifiers use a combination of geometric, kinematic and PID variables to distinguish signal from background. The geometric quantities include the $\chisqip$ of all final-state particles, the three-dimensional and radial flight distances of the $\Lb$ candidate decay vertex from its associated PV, the $\chi^2$ of the $\Lb$ candidate vertex fit, the angle $\theta_{\vec{p},V}$ between the momentum vector of the $\Lb$ candidate and the line that joins the $\Lb$ decay vertex and the PV, the $\chisq$ of the $\Lc$ decay-vertex fit, and the decay time of the $\Lc$ candidate.
The kinematical quantities include the total momentum and transverse momentum of each final-state particle.
For each particle, a probability for the assigned PID hypothesis~\cite{LHCb-DP-2018-001} is also used. The PID response of the charged hadrons in the simulated signal and normalization mode decays is obtained from dedicated calibration samples from the data where no PID requirements are imposed~\cite{LHCb-DP-2012-003,LHCb-DP-2018-001}.
For the $3\pi$ sample, two additional variables, $M(\pim\pip\pim)$ and the $\chisq$ of the vertex separation between the $3\pi$ vertex and the PV, are included.

To train the BDT1 classifiers, signal $\Lb$ decays are taken from simulated $\Xibm\to\Lb\pims$ decays, where the $\Lb$ baryon is forced to decay into either the $\Lc\pim$ or $\Lc\pim\pip\pim$ mode. The latter is simulated with a number of intermediate resonances to reproduce the two and three-body masses observed in the data. The usage of the $\Xibm\to\Lb\pims$ simulation will favor selecting $\Lb$ baryons that are produced in the decay of a $\Xibm$ baryon. The combinatorial background sample for the BDT1 training is composed of $\Lb$ candidates in the data that have invariant mass in the range $5700 < M(\Lc\pim[\pip\pim]) < 5850$\mevcc. 

\begin{figure}[tb]
	\centering
	\includegraphics[width=0.48\textwidth]{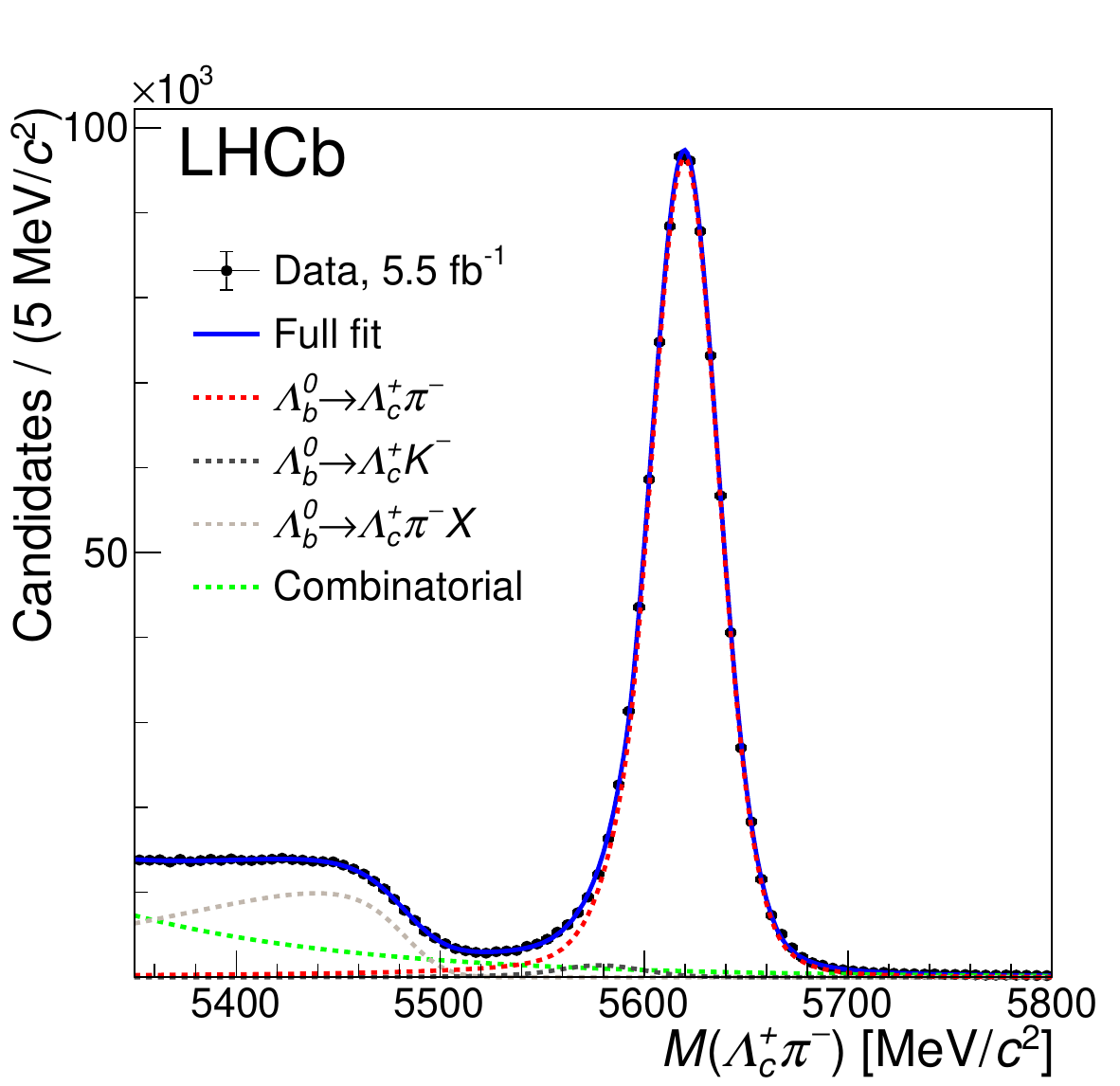}
    \includegraphics[width=0.48\textwidth]{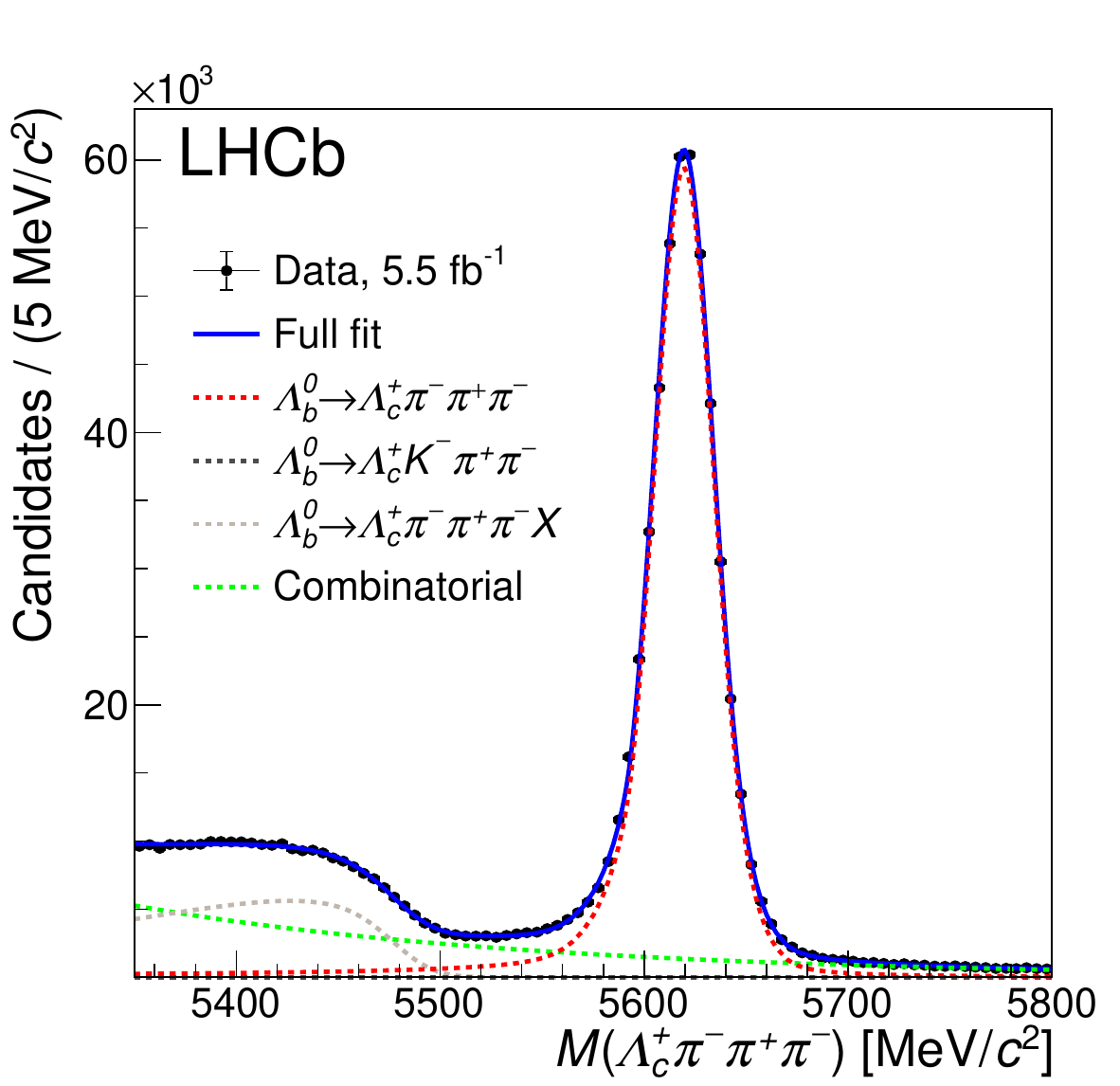}
	\caption{\small{Invariant-mass spectra of (left) \Lb\to\Lc\pim and (right) \Lb\to\Lc\pim\pip\pim candidates for the full data sample. The fitted signal and background shapes, as described in the text, are overlaid.}}
	\label{fig:LbMassFitFull}
\end{figure}

The optimal requirement on the BDT1 response for each $\Lb$ decay mode is determined by taking the maximum value of the figure of merit (FOM), defined as FOM\,$=S/\sqrt{S+B}$, where $S=S_0\eff_{\rm BDT1}$. The estimated number of $\Xibm$ signal decays with no BDT1 requirement, $S_0$, is taken to be 200, based on an extrapolation of the results in Ref.~\cite{LHCb-PAPER-2015-047}, and $\eff_{\rm BDT1}$ is the efficiency of the BDT1 requirement on simulated signal decays, which includes the weights discussed previously.  The quantity $B$ is the expected number of background events in the range $34.8<\delta M < 41.6$\mevcc, about 2.5 times the expected mass resolution. The FOM is found to have a broad and nearly flat maximum with a drop near the endpoints of $-1$ and 1. A loose requirement of ${\rm BDT1}>0$ is chosen for both $\Lb$ modes, resulting in an efficiency of 93\% for selecting $\Lb$ baryons in $\Xibm\to\Lb\pims$ decays and 87\% for promptly produced $\Lb$ baryons, while removing about 80\% of the combinatorial background.

The mass spectra of $\Lb$ candidates passing all selection requirements, except the final $\Lb$ mass window of $5560<M(\Lb)<5680$\mevcc, are shown in Fig.~\ref{fig:LbMassFitFull}. The mass distributions are described as the sum of a signal shape and several background components, and a binned extended maximum-likelihood fit is performed to obtain the $\Lb$ signal yields. The signal shapes are obtained from simulated decays, and are modeled as the sum of two Crystal Ball functions~\cite{Skwarnicki:1986xj} with power law tails on each side and a common peak value. All parameters are fixed to the values estimated from the simulation except for the peak mass value and an overall scale factor for the resolution to account for a small difference between the mass resolution in data and simulation. The invariant mass shapes describing the partially reconstructed ${\Lb\to\Lc\pim(\pip\pim)X}$ decays, where $X$ represents one or more missing particles, and misidentified $\Lb\to\Lc\Km(\pip\pim)$ background contributions are based on simulated decays, and are described in Ref.~\cite{LHCb-PAPER-2014-021}. The combinatorial background is parameterized with an exponential function.  Signal yields of $(921\pm1)\times10^3$ and $(511\pm1)\times10^3$ in the full fit range are obtained in the $\Lb\to\Lc\pim$ and $\Lb\to\Lc\pim\pip\pim$ modes, respectively. The mass resolution scale factor is 1.1, which is consistent with that found in other $b$-hadron decay analyses. The background-to-signal ratios in the signal region from 5560--5680\mevcc are 2.1\% and 6.8\% for the \Lb\to\Lc\pim and \Lb\to\Lc\pim\pip\pim decay modes, respectively.

In forming the $\Xibm$ candidates, the major sources of background are from random combinations of $\Lb$ baryons and $\pim$ mesons, and from the strong decays $\Sigmabsm\to\Lb\pim$. Both of these backgrounds have a candidate decay-time distribution that peaks at zero, whereas the signal has preferentially positive decay times.
Ordinarily, this would be a powerful discriminant against background, but due to the low $\pim$ meson momentum and relatively small opening angle in the $\Xibm$ baryon decay, the uncertainty on the $\Xibm$ vertex position is large. The resolution on the $z$ coordinate of the $\Xibm$ decay vertex is about 7~mm, which is to be compared to about 0.5~mm for the $\Lb$ decay vertex. This relatively large uncertainty makes it more challenging to separate $\Lb\pim$ background from the $\Xibm\to\Lb\pims$ signal.

The second BDT classifier, BDT2, is employed to suppress combinatorial background, primarily from prompt $\Lb\pim$ combinations, and uses 12 discriminating variables: the $\pt$, flight distance, and $\theta_{\vec{p},V}$ of the $\Xibm$ candidate; the decay time, $p$ and $\pt$ of the $\Lb$ baryon; the $p$, $\pt$ and $\chisqip$ of the $\pim$ meson and its associated PID probability to be a pion; and $\eta(\Lb)-\eta(\pim)$ and $\phi(\Lb)-\phi(\pim)$, where $\eta(X)$ and $\phi(X)$ are the pseudorapidity and azimuthal angle of the indicated particles. Simulated signal decays are used to to train the BDT2 classifier. Wrong-sign candidates within $2.5\,\sigma$ of the expected signal peak position in $\delta M$ are used to represent the background. 

A second FOM, defined in an analogous way to that of BDT1, is used to optimize the BDT2 selection requirement. The maximum value of the FOM corresponds to ${\rm BDT2}>0.9$, which is used for the signal significance. For the branching fraction measurement, a looser requirement of ${\rm BDT2}>0.8$ is used, which gives about 30\% higher relative efficiency according to simulation with only a slightly lower FOM. Hereafter, these are referred to as the tight (${\rm BDT2} >0.9$) and loose (${\rm BDT2}>0.8$) selections. The looser selection on BDT2 reduces the systematic uncertainty associated with the BDT2 requirement. The two BDT2 requirements, their efficiencies and FOM values are shown in Table~\ref{tab:BDT}. 

\begin{table*}[tb]
\begin{center}
\caption{\small{Efficiencies for the two BDT2 selections on simulated $\Xibm$ signal decays and on WS background (in \%) for the two $\Lb$ final states. The FOM values are also shown. Uncertainties are statistical only.}}
\begin{tabular}{lcccccc}
\hline\hline
\\[-2.2ex]
BDT2~ & \multicolumn{2}{c}{$\eff_{\rm BDT2}^{\Xibm~~~~~~~}$}  & \multicolumn{2}{c}{$\eff_{\rm BDT2}^{\rm bkg}$~~~~~~~~~~} & \multicolumn{2}{c}{FOM~~~~~~~~~~} \\
selection & & & & & & \\
[-1.2ex]
 & \Lc\pim & \Lc\pim\pip\pim & \Lc\pim & \Lc\pim\pip\pim & \Lc\pim & \Lc\pim\pip\pim \\
 \\[-2.4ex]
\hline
 \\[-2.4ex]
Tight   & $35.6\pm0.2$ & $39.9\pm0.2$ & $0.6\pm0.1$ & $1.1\pm0.1$ & 5.4  & 6.2\\
Loose   & $46.1\pm0.2$ & $51.8\pm0.2$ & $1.4\pm0.1$ & $2.6\pm0.2$ & 5.0 & 5.7 \\
\hline\hline
\end{tabular}
\label{tab:BDT}
\end{center}
\end{table*}

\begin{figure}[tb]
	\centering
	\includegraphics[width=0.49\textwidth]{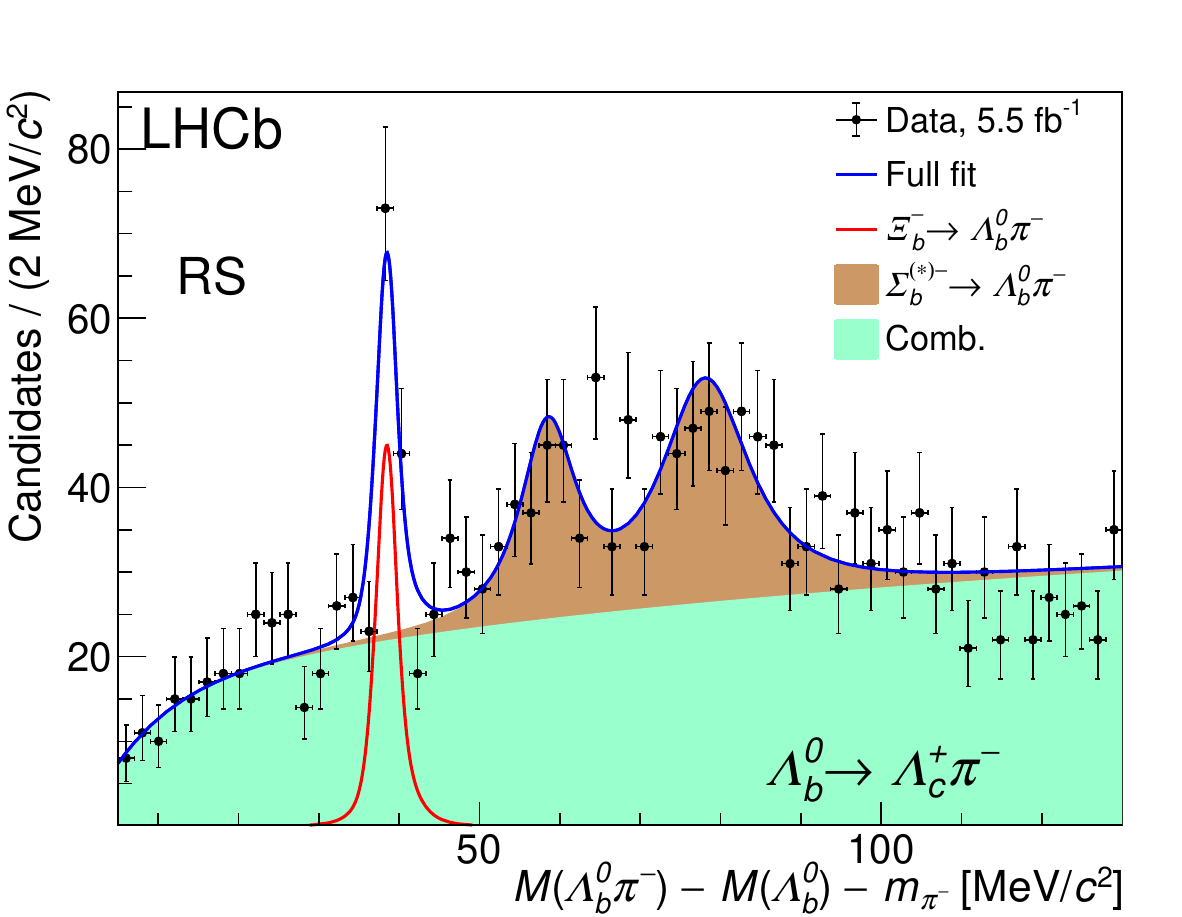}
	\includegraphics[width=0.49\textwidth]{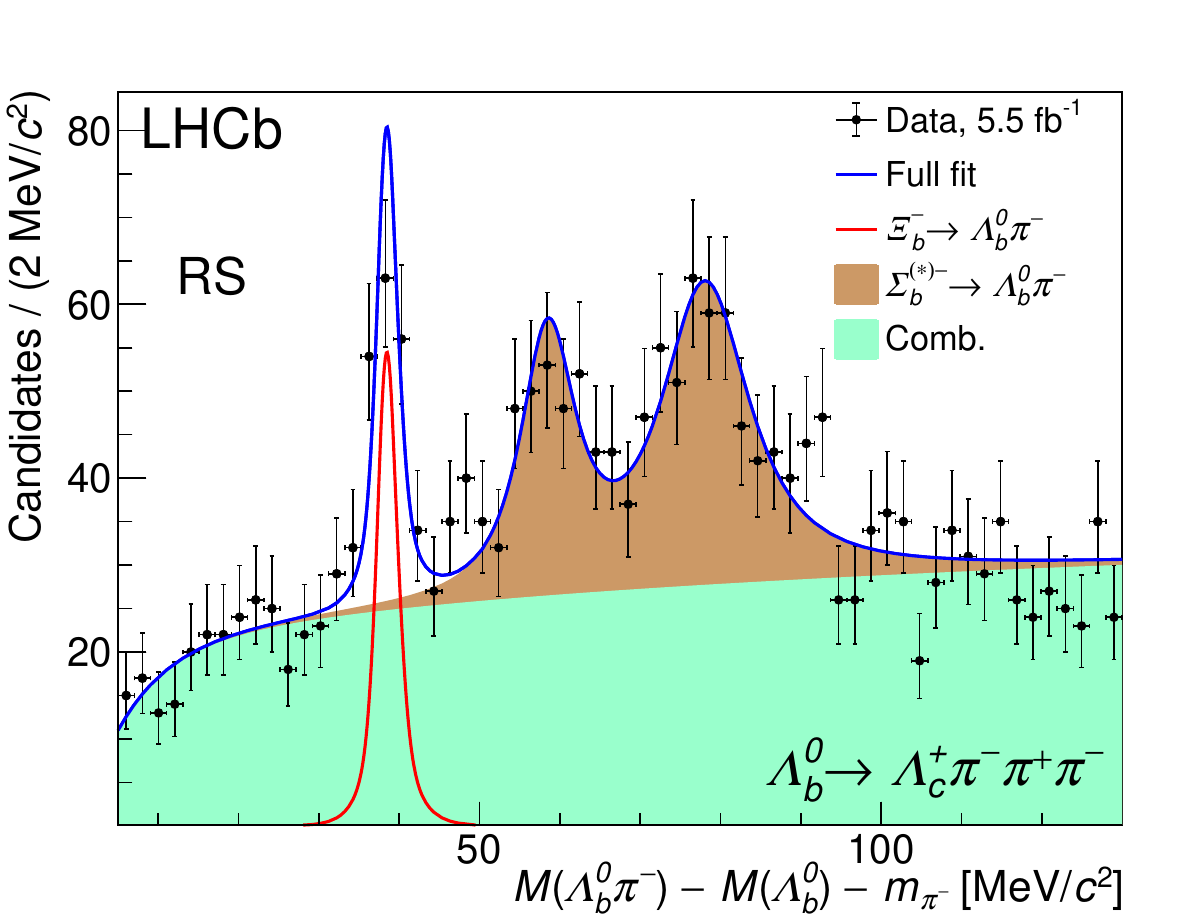}
    \includegraphics[width=0.48\textwidth]{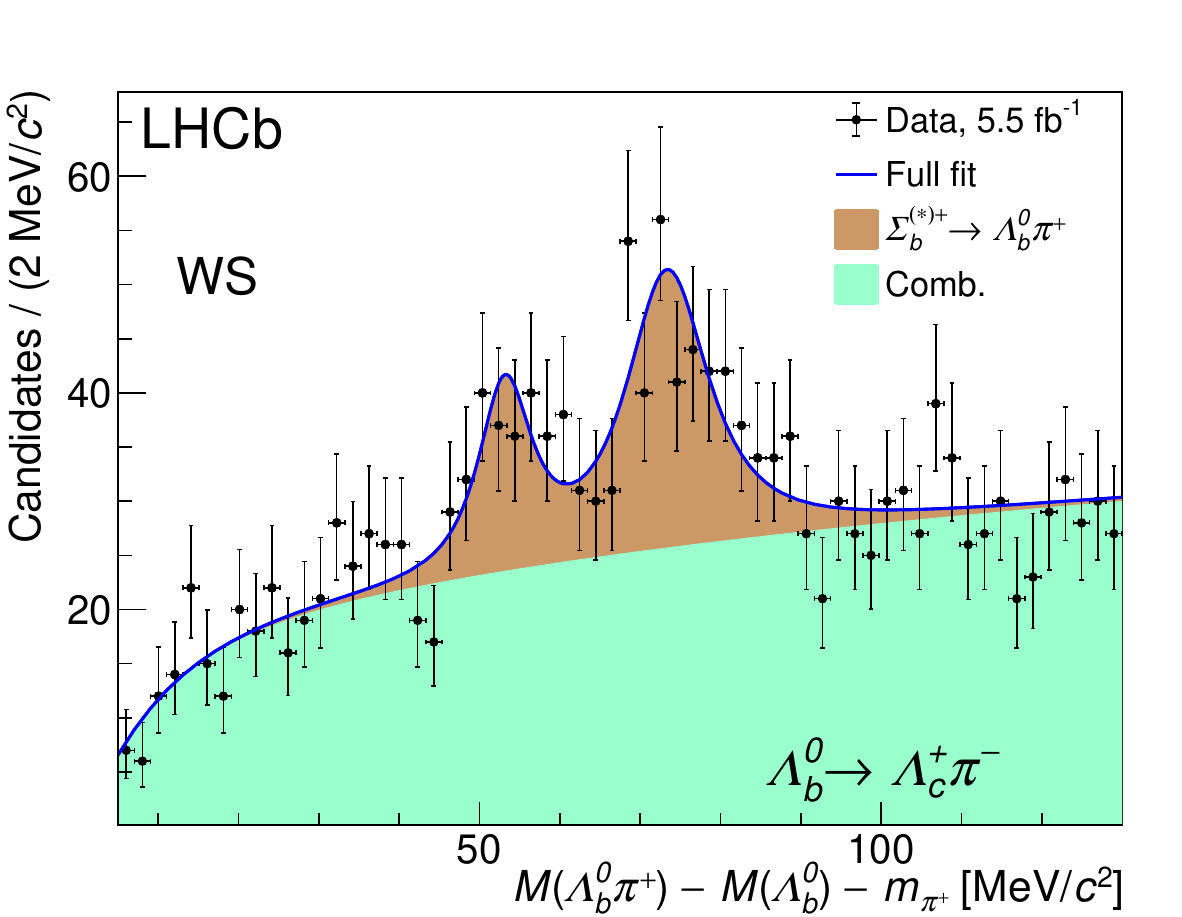}
	\includegraphics[width=0.48\textwidth]{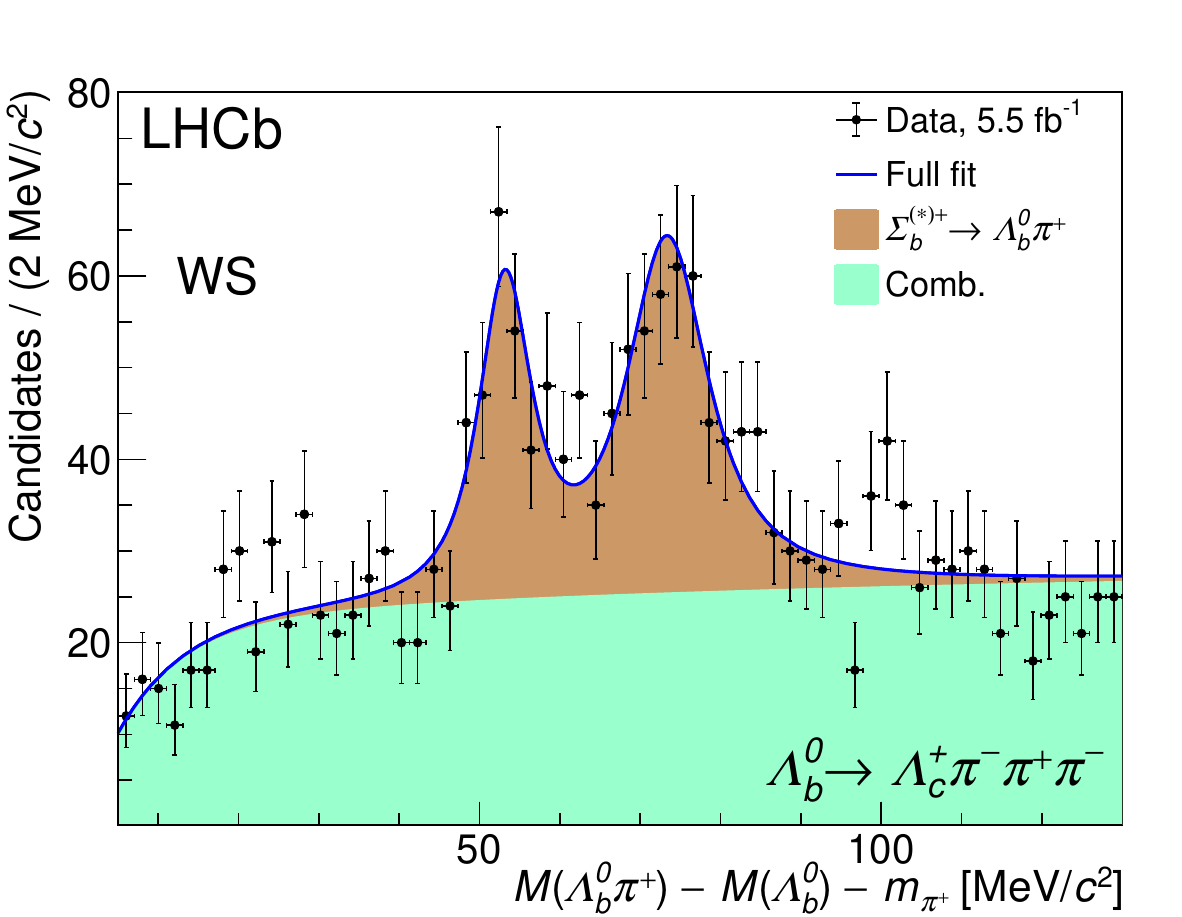}
	\caption{\small{(Top row) Spectra of the mass difference, $M(\Lb\pim)-M(\Lb)-m_{\pim}$ for $\Xibm\to\Lb\pim$ candidates, for the	(left) $\Lb\to\Lc\pim$ and (right) $\Lb\to\Lc\pim\pip\pim$ right-sign samples, with the tight BDT2 selection. The bottom row shows the corresponding distributions for the wrong-sign candidates. Fits to the data are overlaid as described in the text.
	}}
	\label{fig:XibMassFit_tight}
\end{figure}

\begin{figure}[tb]
	\centering
	\includegraphics[width=0.49\textwidth]{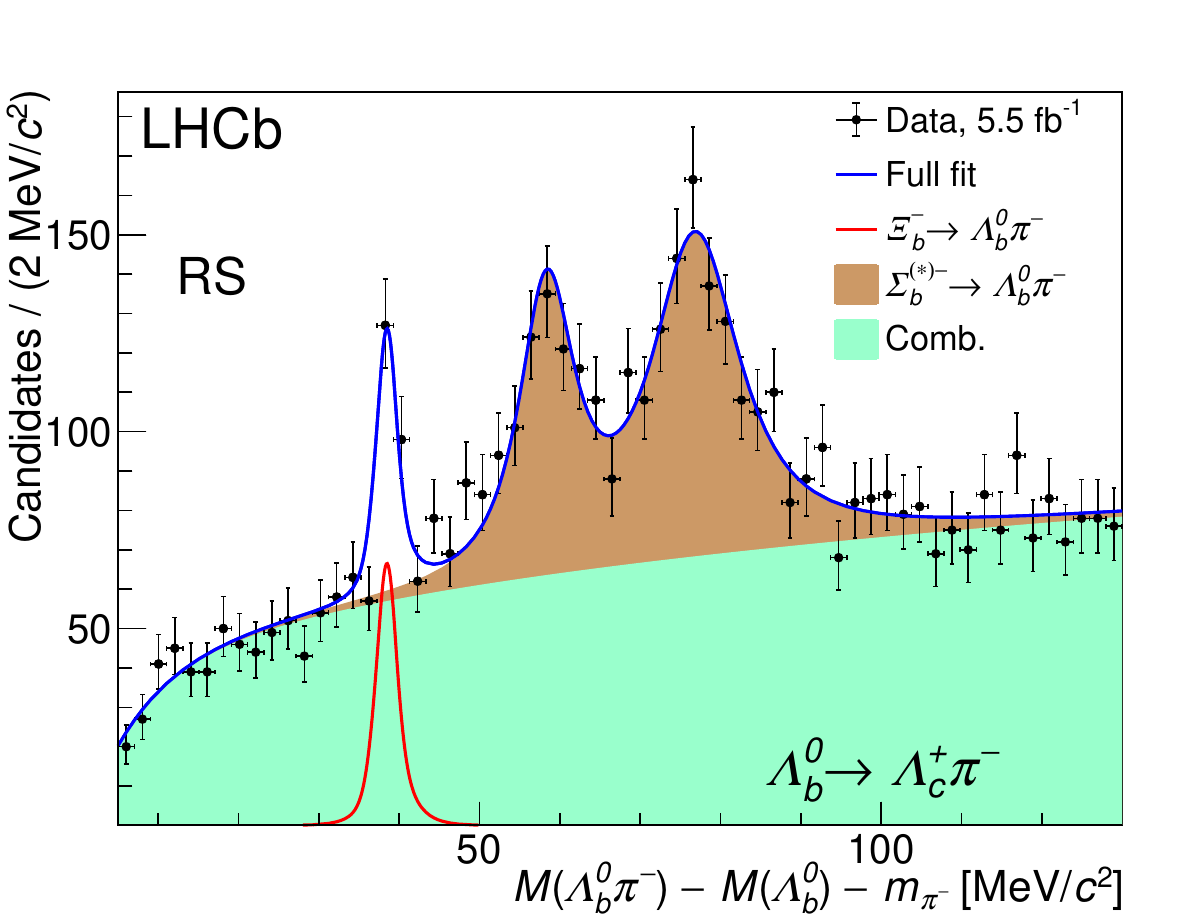}
	\includegraphics[width=0.49\textwidth]{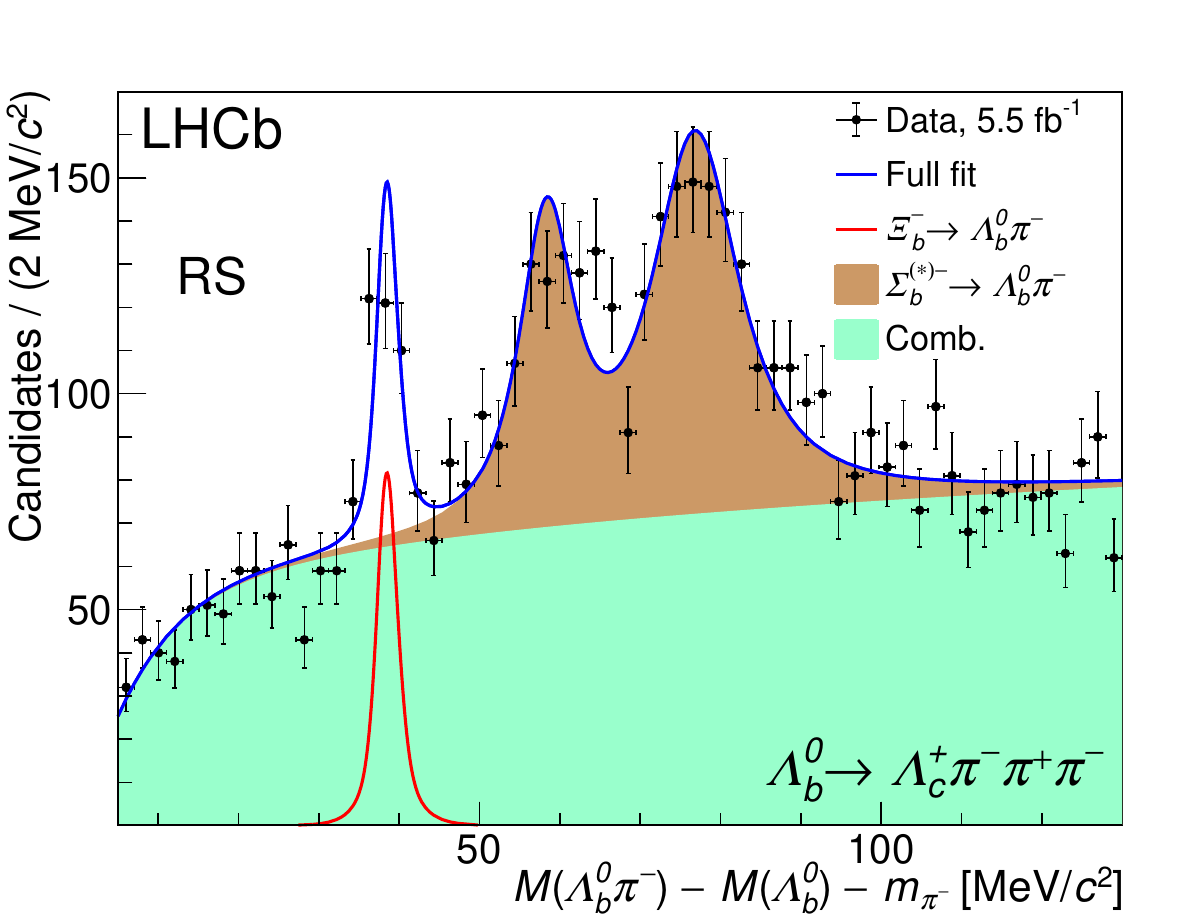}
    \includegraphics[width=0.49\textwidth]{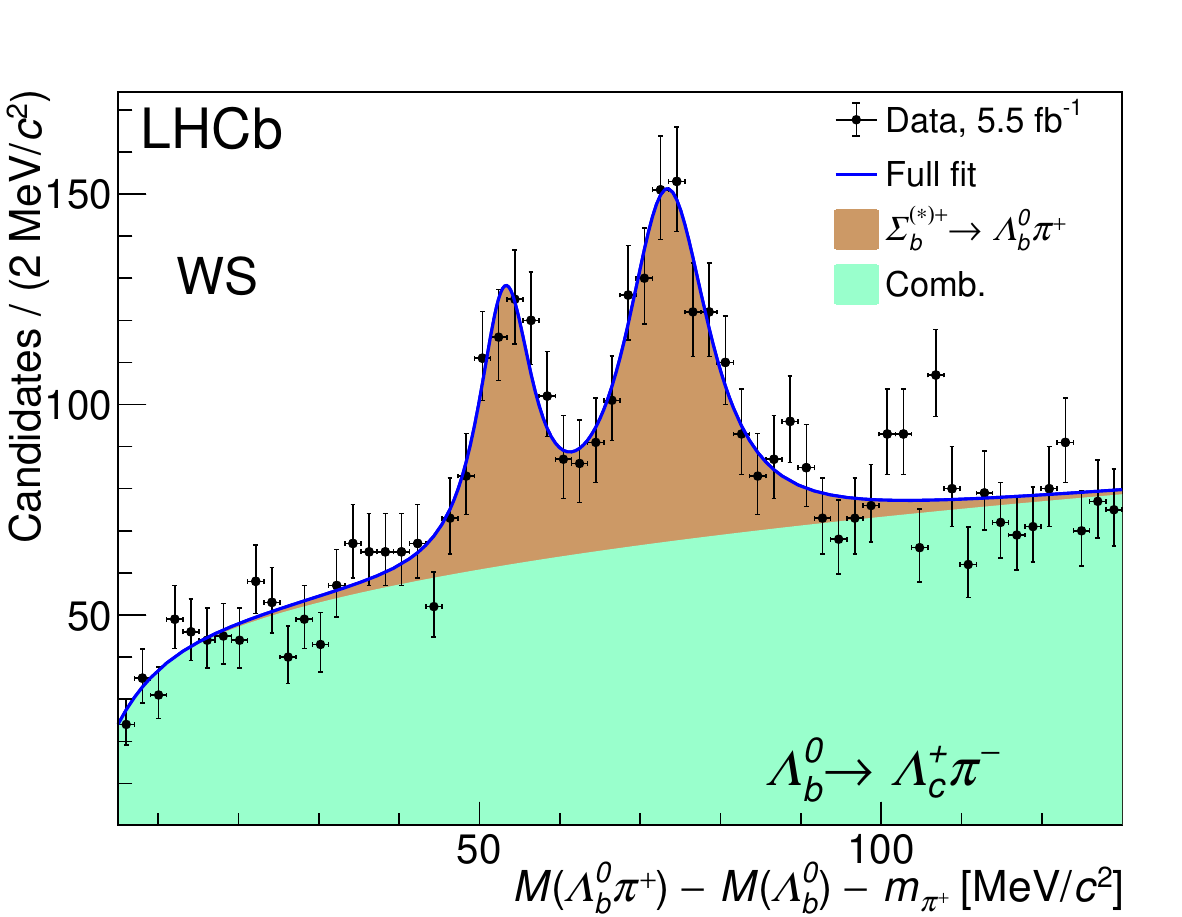}
	\includegraphics[width=0.49\textwidth]{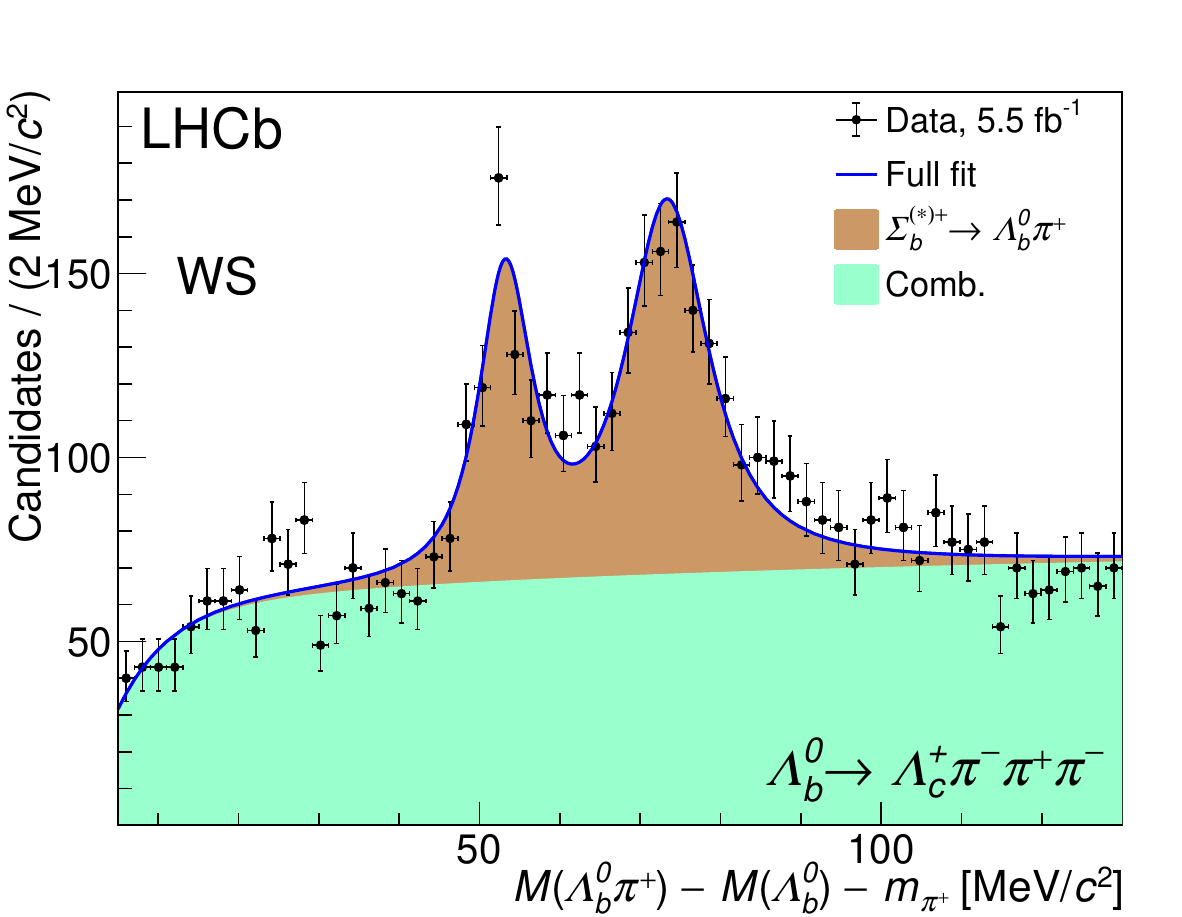}
	\caption{\small{(Top row) Spectra of the mass difference, $M(\Lb\pim)-M(\Lb)-m_{\pim}$ for $\Xibm\to\Lb\pim$ candidates, for
    the	(left) $\Lb\to\Lc\pim$ and (right) $\Lb\to\Lc\pim\pip\pim$ right-sign samples with the loose BDT2 selection. The bottom row shows the corresponding distributions for the wrong-sign candidates. Fits to the data are overlaid as described in the text.
	}}
	\label{fig:XibMassFit}
\end{figure}

Figure~\ref{fig:XibMassFit_tight} shows the $\delta M$ spectra for the selected $\Xibm\to\Lb\pim$ signal candidates passing the tight BDT2 requirement for the $\Lb\to\Lc\pim$ and $\Lb\to\Lc\pim\pip\pim$ samples. The corresponding distributions with the loose BDT2 selection for the $r_s$ measurement are shown in Fig.~\ref{fig:XibMassFit}. In both cases, there are clear peaks at the expected location for a $\Xibm$ signal in the RS spectra, and no such peak in the WS spectra.

A simultaneous unbinned extended maximum likelihood fit to the four (RS and WS for each \Lb mode) $\delta M$ spectra is performed to obtain the $\Xibm$ signal yields. The spectra are modeled as the sum of a $\Xibm$ signal shape, and background shapes for the $\Sigmabspm$ resonances and random $\Lb\pim$ combinations. The $\Xibm$ signal shape for each $\Lb$ mode are obtained from simulated signal decays, and is described by the sum of two Crystal Ball functions~\cite{Skwarnicki:1986xj}. All shape parameters are fixed to the values obtained from simulation except for the peak mass value. When fitting the data, the mass resolution parameters in the Crystal Ball function are scaled by 1.1 to account for the larger mass resolution in data than in simulation. The $\Sigmabspm$ resonances are modeled by a relativistic Breit--Wigner function, as described in Ref.~\cite{LHCb-PAPER-2015-047}, convolved with the mass resolution. The peak mass values of the $\Sigmabspm$ signals are biased by about $+$2\mevcc due to the selection favoring those decays that are reconstructed with positive displacement relative to the PV, and for this reason, the peak mass values are allowed to vary freely in the fit. The widths are fixed to the world average values~\cite{PDG2022}. The combinatorial background is described by a threshold function of the form $(\delta M)^A(1-e^{-\delta M/C})$. The parameters $A$ and $C$ are freely varied in the loose BDT fit, but the low background level in the tight BDT fit is insufficient to constrain the parameter $A$. It is therefore fixed in the tight BDT fit to the value obtained in the loose BDT fit. Among the freely varied parameters in the fit, only the peak mass values of the $\Xibm$ signal and the $\Sigmabspm$ resonances are shared between the two $\Lb$ modes.

\begin{figure}[htb]
	\centering
	\includegraphics[width=0.49\textwidth]{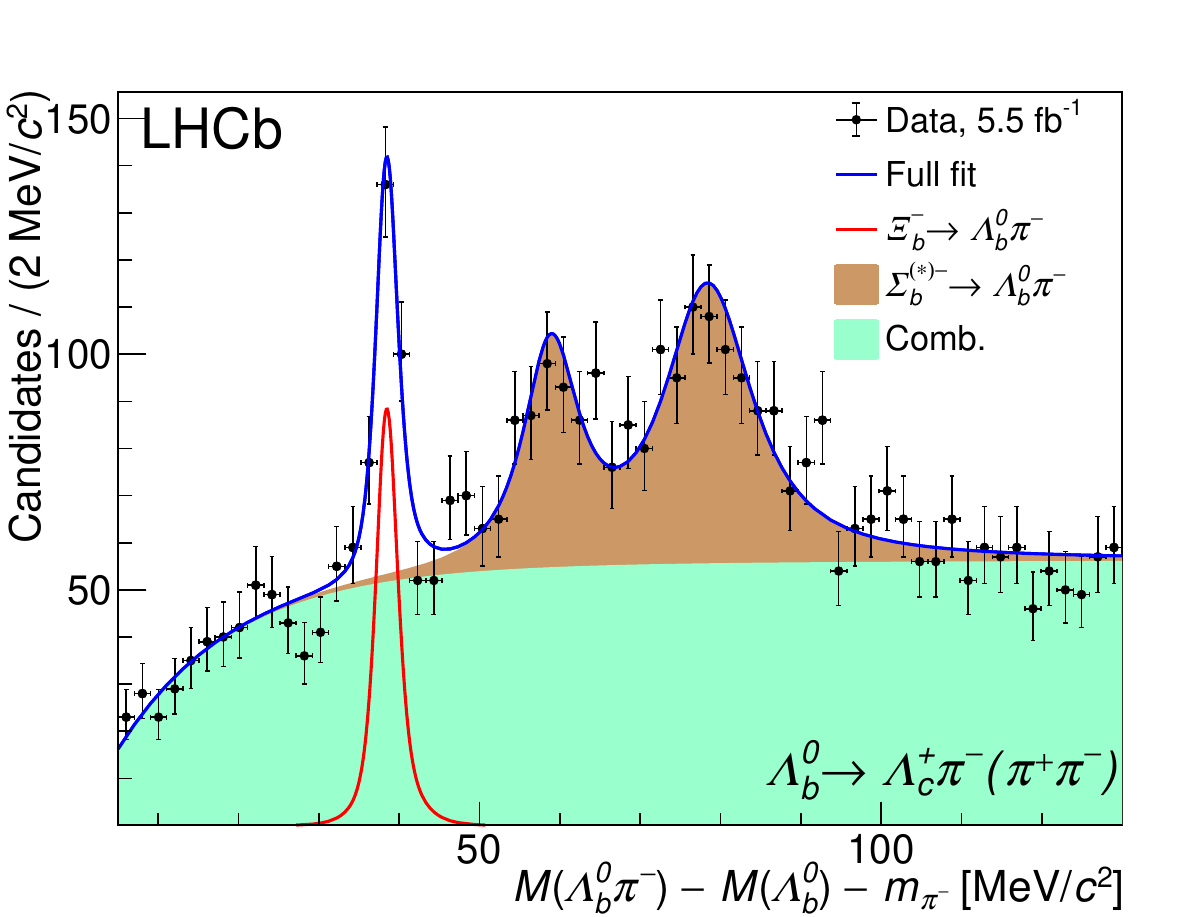}
  	\includegraphics[width=0.49\textwidth]{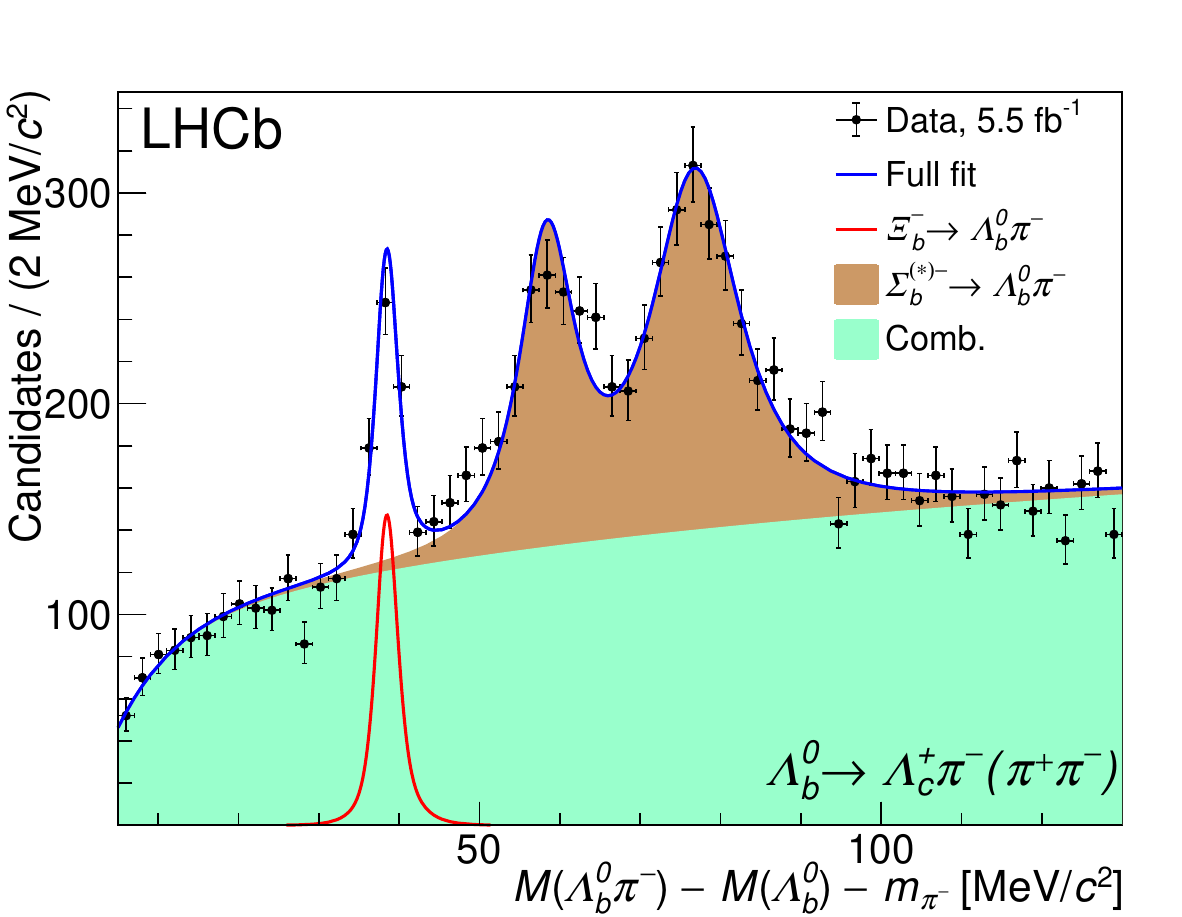}
	\caption{\small{Spectra of the mass difference, $M(\Lb\pim)-M(\Lb)-m_{\pim}$ for $\Xibm\to\Lb\pim$ candidates for the combined $\Lb\to\Lc\pim$ and $\Lb\to\Lc\pim\pip\pim$ samples for (left) tight BDT2 selection and (right) loose BDT2 selection.}}
	\label{fig:XibMassFitBothTight}
\end{figure}

The results of the fits are overlaid in Figs~\ref{fig:XibMassFit_tight} and~\ref{fig:XibMassFit}. The $\Xibm$ signal yields with the tight BDT2 selection, used only for maximizing the signal significance, are $85\pm13$ and $103\pm15$ for the $\Lb\to\Lc\pim$ and $\Lb\to\Lc\pim\pip\pim$ modes, respectively. Using Wilks theorem~\cite{Wilks:1938dza}, the significance of the pair of $\Xibm$ signal peaks corresponds to about 11$\sigma$, thus definitively establishing the first observation of the $\Xibm\to\Lb\pim$ decay. For the loose BDT2 selection, the corresponding $\Xibm$ signal yields are $126\pm19$ and $154\pm23$. The larger yield in the $3\pi$ mode is anticipated, based upon the larger relative efficiency (discussed later) and the measured yields of $\Lb$ decays. The $\Xibm$ peak position in $\delta M$ is at $38.5\pm0.2\stat$\mevcc, which is about 1$\sigma$ from the value of $38.14\pm0.29$\mevcc, based on a recent precise measurement of the $\Xibm$ mass using $\Xibm\to\Xicz\pim$ decays~\cite{LHCb-PAPER-2020-032}. No narrow peaks are seen in the WS spectra. Corresponding fits to the combined sample, where the fraction of $\Xibm$ signal in the $\Lb\to\Lc\pim$ mode is fixed to 45\% of the total, are shown in Fig.~\ref{fig:XibMassFitBothTight}.

\section{Selection efficiencies}
Simulated decays of the signal and normalization modes are used to estimate the relative efficiency, $\eff_{\rm rel}=\eff_{\rm sig}/\eff_{\rm norm}$, where $\eff_{\rm sig}$ ($\eff_{\rm norm}$) 
refers to the efficiency to reconstruct and select the $\Xibm$ signal ($\Lb$ normalization) mode. For the normalization mode efficiency, production of $\Lb$ baryons through weak decays, which contribute at the $10^{-3}$ level, are neglected. Relative efficiencies are determined for each of the $\Lb$ decay modes, $\Lc\pim$ and $\Lc\pim\pip\pim$. The relative efficiencies are computed within the $\pt<20$\gevc, $2<\eta<6$ fiducial acceptance. Since the $\Lb$ decay is common to the signal and normalization mode, and the low momentum pion from the $\Xibm$ decay contributes negligibly to the hardware or software trigger, no specific selections are made on the trigger decisions. 

To account for small differences between the simulation and data, a multiplicative correction to the relative efficiency is applied, defined as
\begin{align}
    \eff_{\rm rel}^{\rm corr}=\eff_{\rm rel}^{\rm track}\,\eff_{\rm rel}^{\rm BDT2}\,\eff_{\rm rel}^{\rm trig}.
\end{align}
The first correction accounts for small differences in the tracking efficiency, and is measured using large samples of $\jpsi\to\mup\mun$ calibration samples~\cite{LHCb-DP-2013-002}. A correction factor ${\eff_{\rm rel}^{\rm track}=0.99\pm0.03}$ is obtained.

The source of the second correction is the BDT2 requirement. The simulation indicates an efficiency of about 50\% for the loose BDT2 selection. To probe whether the BDT2 distribution in data and simulation are compatible, the signal yield ratio, $r_{\rm BDT2}$, for the loose to tight BDT2 selection is compared. From the fitted yields in the two cases, the ratios are $r_{\rm BDT2}^{1\pi}=1.48\pm0.17$ and $r_{\rm BDT2}^{3\pi}=1.50\pm0.18$, where the expected value using efficiencies from simulation is 1.30 for both $\Lb$ decay modes. Although both ratios in data are within about one standard deviation of the expected value, they are both larger, suggesting that the BDT2 response from simulation may be more strongly peaked at unity than the data. This supposition is supported by comparing the BDT1 distributions in simulation and background-subtracted data, where the large $\Lb$ sample in simulation is seen to be more sharply peaked at unity than the data. A correction is derived by smearing the BDT2 distribution from simulation with a one-sided Gaussian function in order to reproduce the $r_{\rm BDT2}$ values obtained in data. The values of $\eff_{\rm rel}^{\rm BDT2}$ obtained are  $0.94\pm0.03$ and $0.93\pm0.04$ for the $1\pi$ and $3\pi$ modes, respectively. 

The third correction accounts for a slightly different hardware trigger efficiency between data and simulation. Two classes of events are studied: (1) those events where the signal decay products result in the event passing the hardware trigger (TOS, short for Triggered on Signal), and (2) cases where other activity in the event, such as the other $b$ hadron or the beam fragments, produces the hardware trigger (TIS, short for Triggered Independently of the Signal). Both are studied  in data and simulation using the TISTOS method~\cite{LHCb-DP-2012-004}. A given event can pass the TOS, TIS, or both the TOS and TIS requirements. Briefly, by selecting TIS events, one can measure the efficiency of TOS events, and vice versa. The benefit of this method is that it can be applied identically to both data and simulation. Using this method, the relative trigger efficiency between data and simulation, $\eff_{\rm rel}^{\rm trig}$, is found to be $1.033\pm0.017$ and $1.008\pm0.004$ for the $1\pi$ and $3\pi$ samples, respectively.

The products of the three correction factors, ${\eff_{\rm rel}^{\rm corr}}$, are $0.972\pm0.046$ and $0.941\pm0.046$ for the $\Lb\to\Lc\pim$ and $\Lb\to\Lc\pim\pip\pim$ decay modes, respectively. For the loose BDT2 requirement, the relative efficiencies, including the above corrections, are found to be
\begin{align}
\eff_{\rm rel}^{\Lc\pim}&=0.210\pm0.004\stat,\\
\eff_{\rm rel}^{\Lc\pim\pip\pim}&=0.394\pm0.008\stat. 
\end{align}
A significantly larger relative efficiency is obtained when the $\Xibm$ signal decay is reconstructed in the $\Lb\to\Lc\pim\pip\pim$ mode as compared to the $\Lc\pim$ mode.
There are two main factors that contribute to this enhancement. The first arises from the $\chisqip$-related selection requirements in the trigger and analysis-related selections. In the $\Lc\pim$ mode, all final-state particles tend to have fairly high $\pt$, and thus large values of $\chisqip$. The larger decay distance of the $\Lb$ baryon from the PV in the signal mode leads to only moderate increases in $\eff_{\rm sig}$ relative to $\eff_{\rm norm}$. However, for the $\Lc\pim\pip\pim$ final state, the 3 pions tend to have lower $\pt$, leading to smaller values of $\chisqip$. If any of the pions have $\chisqip<4$, the $\Lb$ candidate is not selected. 
The combination of (1) the larger displacement of the $\Lb$ baryon from the PV (due to the $\Xibm$ baryon lifetime) for the signal mode as compared to the normalization mode and (2) the lower $\pt$ of the pions, leads to a significantly larger increase in $\eff_{\rm sig}$ relative to $\eff_{\rm norm}$ for the $\Lc\pim\pip\pim$ final state. The second effect that contributes to the increase in $\eff_{\rm rel}$ is the difference in the average $p$ and $\pt$ of reconstructed $\Lb$ decays for each of the two final states. Higher multiplicity final states must, on average, have larger momentum in order to be reconstructed and pass all selection requirements. This requires a higher momentum $\Xibm$ baryon when reconstructed in the $3\pi$ mode than the $1\pi$ mode, which in turn leads to a higher relative efficiency for reconstructing the $\pim$ from the $\Xibm$ decay.

A summary of the yields in the signal and normalization modes, the relative efficiencies, and the computed $r_s$ values are shown in Table~\ref{tab:Yields}. The $\Lb$ yields include about a 1\% contribution from misidentified $\Lb\to\Lc\Km(\pip\pim)$ decays, which also have a narrow peak in $\delta M$ at about 38\mevcc. The $r_s$ values for the two $\Lb$ modes are statistically compatible with each other. 

\begin{table*}[tb]
\begin{center}
\caption{\small{Summary of the yields in the signal ($N(\Xibm)$) and normalization modes ($N(\Lb)$), the relative efficiencies ($\eff_{\rm rel}$), and the resulting $r_s$ values for the loose BDT selection. The $\Lb$ signal yields are in the invariant-mass region $5560<M(\Lb)<5680$\mevcc. Uncertainties are statistical only.}} 
\begin{tabular}{lcc}
\hline\hline
Fit parameter  & ~~\Lc\pim & \Lc\pim\pip\pim  \\
  \\[-2.5ex]
\hline
  \\[-2.5ex]
$N(\Xibm\to\Lb\pim)$   & $~\,126\pm19$   & $~\,154\pm23$      \\
$N(\Lb)~[10^3]$   & $879\pm1$   & $483\pm1$    \\
$\eff_{\rm rel}~[10^{-2}]$ &  $~21.0\pm0.4$ & $~\,39.4\pm0.8$  \\
$r_s~[10^{-4}]$ & $~~\,6.80\pm1.05$  & $~~~8.09\pm1.23$  \\
\hline\hline
\end{tabular}
\label{tab:Yields}
\end{center}
\end{table*}

\begin{table*}[tb]
\begin{center}
\caption{\small{Systematic uncertainties in the measurement of $r_s$ for each of the two decay modes of the $\Lb$ baryon. The horizontal dividing line separates the uncorrelated (above) and correlated (below) uncertainties}. For the latter, the listed sources are 100\% correlated between the two $\Lb$ decay modes.}
\begin{tabular}{lcc}
\hline\hline
Source           & \multicolumn{2}{c}{Value (\%)} \\
                 & $\Lc\pim$ & $\Lc\pim\pip\pim$ \\
\hline
$\Xibm$ signal shape     &       1.4      &  2.4   \\
$\Xibm$ background shape  &      3.1      &  1.8 \\
$\Lb$ signal shape       &       0.3      &  0.8    \\
$\Lb$ background shape   &       0.1     & 0.7      \\
Geom. acceptance          &      1.8      &  1.8    \\
Sim. weights \& sample sizes  &  3.6      &  3.4     \\
Trigger efficiency       &       1.7      &  0.4    \\
\hline
$\Xibm$ $\pt$ spectrum   &       3.2      &  5.6     \\
IP resolution            &       1.3      &  0.7      \\
BDT2 efficiency          &       3.0      &  3.5     \\
Tracking efficiency        &       3.3      &  3.3    \\
Multiple candidates      &       0.5      &  2.6    \\
$\Xibm$ lifetime         &       3.0      &  2.5    \\       
\hline
Total                    &       8.5      &  9.7     \\
\hline\hline
\end{tabular}
\label{tab:Syst}
\end{center}
\end{table*}

\section{Systematic uncertainties}
Systematic uncertainties on $r_s$ are summarized in Table~\ref{tab:Syst}, and have contributions related to the determination of the signal and normalization mode yields and the relative efficiencies. The systematic uncertainty in the $\Xibm$ signal yields is obtained from the fractional difference in the yields when increasing and decreasing the resolution scale factor by $\pm$5\%. For the normalization modes, different bin widths and fits allowing all $\Lb$ shape parameters to vary freely are performed. The bin widths have negligible effect on the signal yields. The change in the $\Lb$ yield between the alternative and baseline fit is assigned as the systematic uncertainty. The $\Lb$ background shape uncertainty is taken as the fractional change in yield when using a second-order Chebychev polynomial in place of the baseline shape. For the $\Xibm$ combinatorial background, the fractional change in yield between the baseline shape and the alternative three-parameter function ${\mathcal{P}}(\delta M)=f(1-e^{-\delta M/A})+(1-f)(1-e^{-\delta M/B})$ is assigned as the systematic uncertainty. 

For the relative efficiency, a number of systematic uncertainties are considered.
Those uncertainties associated with the selections on the $\Lb$ candidates are common to both the signal and normalization mode, and are significantly reduced. 

The uncertainty in the geometric acceptance is due to the usage of finite-sized samples of simulated decays. The weights applied to the simulated decays to replicate the $(\pt,\eta)$ distributions in data are limited by the finite yields of $\Lb\to\Lc\pim(\pip\pim)$ and $\Xibm\to\Xicz\pim$ decays in simulation and in the data. The uncertainty is assigned by varying all of the weights within their uncertainties, and recomputing the relative efficiency. This procedure is carried out 100 times, and the standard deviation of the $\eff_{\rm rel}$ distribution is assigned as the uncertainty. 

The $\Xibm$ efficiencies are obtained from simulated signal decays, where the signal weights are obtained from $\Xibm\to\Xiz\pim$ decays. Because of differing kinematics between the control and $\Xibm\to\Lb\pim$ modes, the assigned weights could have some small biases. To estimate the potential magnitude of this bias, the relative efficiencies are re-evaluated using the $\Lb$ weights for the signal mode in place of the $\Xibm$ weights. The fractional change in the relative efficiency is assigned as the systematic uncertainty.

The IP resolution shows some small differences between data and simulation. To estimate how the difference impacts the relative efficiency, a scaling of the $\log(\chisqip)$ values of each track is performed that brings the $\chisqip$ distributions in data and simulation into good agreement. With the scaling, some of the final-state tracks will fail the $\chisqip>4$ requirement, leading to a reduction in the selection efficiency. The procedure is applied to both the signal and normalization modes, and the fractional change in the relative efficiency is assigned as the systematic uncertainty.

As discussed previously, the BDT2 distribution in simulation is smeared to obtain an efficiency correction for the loose BDT2 requirement, and the uncertainty is taken to be 50\% of the difference of the correction factor from unity. Similarly, the uncertainty in the relative trigger efficiency is assigned to be half of the applied correction.

The tracking efficiency is calibrated using large samples of $\jpsi\to\mup\mun$ decays, and provide a data-to-simulation correction for tracks in the momentum range from 5--200\gevc and $1.9<\eta<4.9$~\cite{LHCb-DP-2013-002}. The corrections are mostly within (1--2)\% of unity. About 65\% of the $\pim$ mesons from the $\Xibm$ decay have momentum below 5\gevc, and for these cases, the correction at 5\gevc is used with an uncertainty that is inflated by a factor of two. The luminosity-weighted average correction is $0.99\pm0.03$. An additional 1.4\% uncertainty is assigned due to a potential difference in the number of hadronic interaction lengths in the simulated and actual detector. 

A possible bias due to keeping all candidates in an event has been studied by comparing the average number of candidates in the $\delta M$ signal region ($34.8<\delta M < 41.6$\mevcc) and in the lower mass sideband region ($27.8<\delta M < 34.6$\mevcc) in data. To the extent that the average number of candidates is the same in these two regions, there is no bias, as it corresponds to an overall increase in the background level. The average number of candidates in the signal region is 1.0 (1.005) for the signal (sideband) regions for the 1$\pi$ mode. The corresponding numbers are 1.046 (1.02) for the 3$\pi$ mode. The difference in the average multiple candidate rate between the signal and sideband regions is assigned as a systematic uncertainty.

The uncertainty in the relative efficiency due to the limited knowledge of the $\Xibm$ lifetime~\cite{PDG2022} is estimated by weighting the simulated $\Xibm$ decay time to produce a smaller lifetime (1.53~ps) and a larger lifetime (1.61~ps), corresponding to a $\pm 1\,\sigma$ variation . The relative change in the $\Xibm$ efficiency is assigned as a systematic uncertainty.

\section{Results and summary}
The two $r_s$ values obtained are averaged, taking the first seven systematic uncertainties in Table~\ref{tab:Syst} as uncorrelated, and the remaining six uncertainties as 100\% correlated. The resulting value is
\begin{align*}
\frac{f_{\Xibm}}{f_{\Lb}}\BR(\Xibm\to\Lb\pim) = (7.3\pm0.8\pm0.6)\times10^{-4},
\end{align*}
where the uncertainties are statistical and total systematic, respectively. Using the independent measurement  
$\frac{f_{\Xibm}}{f_{\Lb}}=(8.2\pm0.7\pm0.6\pm2.5)$\%~\cite{LHCb-PAPER-2018-047}, the branching fraction is determined to be
\begin{align*}
\BR(\Xibm\to\Lb\pim) = (0.89\pm 0.10\pm 0.07\pm0.29)\%,
\end{align*}
\noindent where the last uncertainty is obtained from the quadrature sum of the uncertainties in $\frac{f_{\Xibm}}{f_{\Lb}}$. The corresponding value obtained from the previous measurements of $r_s$~\cite{LHCb-PAPER-2015-047} and $\frac{f_{\Xibm}}{f_{\Lb}}$~\cite{LHCb-PAPER-2018-047} at 7 and 8\tev is readily computed to be ${\BR(\Xibm\to\Lb\pim) = (0.85\pm0.27\pm0.13\pm0.26)\%}$. The measurement reported here has about three times better statistical precision.

In summary, using a $pp$ collision data sample at center-of-mass energy 13~TeV collected by the LHCb experiment, corresponding to an integrated luminosity of 5.5\invfb, the $\Xibm\to\Lb\pim$ decay, in which the $b$ quark is a spectator in the decay process, is observed for the first time. The branching fraction does not show any large enhancements of up to 8\% as suggested in Refs.~\cite{PhysRevD.90.033016,Voloshin:2000et}. The measured branching fraction is consistent with the diquark model calculation of 0.69\% in Ref.~\cite{Cheng2016}, and predictions of (0.19--0.76)\% based on current algebra approaches~\cite{FALLER2015653,PhysRevD.105.094011,PhysRevD.106.093005}, and $(0.63\pm0.42)\%$ based on duality~\cite{GronauPhysRevD.93.034020}. The lower predicted values of $(0.14\pm0.07)\%$ using a non-relativistic constituent quark model~\cite{PenyYuNiuPLB826} or 0.2\% obtained with the MIT bag model~\cite{Cheng2016} are disfavored. The branching fraction obtained here is also compatible with the na\"{\i}ve ratio of total decay widths of the $\Lz$ to $\Lb$ baryons, which is 0.58\%. Although there is no sizeable enhancement in this decay mode, this extra contribution to the $\Xibm$ decay width should be taken into account when comparing the measured $\Xibm$ lifetime to theoretical predictions that only consider the decay of the $b$ quark. 

\section*{Acknowledgements}
%
%
\noindent We express our gratitude to our colleagues in the CERN
accelerator departments for the excellent performance of the LHC. We
thank the technical and administrative staff at the LHCb
institutes.
We acknowledge support from CERN and from the national agencies:
CAPES, CNPq, FAPERJ and FINEP (Brazil); 
MOST and NSFC (China); 
CNRS/IN2P3 (France); 
BMBF, DFG and MPG (Germany); 
INFN (Italy); 
NWO (Netherlands); 
MNiSW and NCN (Poland); 
MCID/IFA (Romania); 
MICINN (Spain); 
SNSF and SER (Switzerland); 
NASU (Ukraine); 
STFC (United Kingdom); 
DOE NP and NSF (USA).
We acknowledge the computing resources that are provided by CERN, IN2P3
(France), KIT and DESY (Germany), INFN (Italy), SURF (Netherlands),
PIC (Spain), GridPP (United Kingdom), 
CSCS (Switzerland), IFIN-HH (Romania), CBPF (Brazil),
Polish WLCG  (Poland) and NERSC (USA).
We are indebted to the communities behind the multiple open-source
software packages on which we depend.
Individual groups or members have received support from
ARC and ARDC (Australia);
Minciencias (Colombia);
AvH Foundation (Germany);
EPLANET, Marie Sk\l{}odowska-Curie Actions, ERC and NextGenerationEU (European Union);
A*MIDEX, ANR, IPhU and Labex P2IO, and R\'{e}gion Auvergne-Rh\^{o}ne-Alpes (France);
Key Research Program of Frontier Sciences of CAS, CAS PIFI, CAS CCEPP, 
Fundamental Research Funds for the Central Universities, 
and Sci. \& Tech. Program of Guangzhou (China);
GVA, XuntaGal, GENCAT, Inditex, InTalent and Prog.~Atracci\'on Talento, CM (Spain);
SRC (Sweden);
the Leverhulme Trust, the Royal Society
 and UKRI (United Kingdom).




\newpage
\addcontentsline{toc}{section}{References}
\bibliographystyle{LHCb}
\bibliography{main,standard,LHCb-PAPER,LHCb-CONF,LHCb-DP,LHCb-TDR}

\ifx\mcitethebibliography\mciteundefinedmacro
\PackageError{LHCb.bst}{mciteplus.sty has not been loaded}
{This bibstyle requires the use of the mciteplus package.}\fi
\providecommand{\href}[2]{#2}
\begin{mcitethebibliography}{10}
\mciteSetBstSublistMode{n}
\mciteSetBstMaxWidthForm{subitem}{\alph{mcitesubitemcount})}
\mciteSetBstSublistLabelBeginEnd{\mcitemaxwidthsubitemform\space}
{\relax}{\relax}

\bibitem{GellMann:1964nj}
M.~Gell-Mann, \ifthenelse{\boolean{articletitles}}{\emph{{A schematic model of
  baryons and mesons}},
  }{}\href{https://doi.org/10.1016/S0031-9163(64)92001-3}{Phys.\ Lett.\
  \textbf{8} (1964) 214}\relax
\mciteBstWouldAddEndPuncttrue
\mciteSetBstMidEndSepPunct{\mcitedefaultmidpunct}
{\mcitedefaultendpunct}{\mcitedefaultseppunct}\relax
\EndOfBibitem
\bibitem{Zweig:352337}
G.~Zweig, \ifthenelse{\boolean{articletitles}}{\emph{{An SU$_3$ model for
  strong interaction symmetry and its breaking; Version 1}}}{}
  \href{http://cds.cern.ch/record/352337}{CERN-TH-401}, CERN, Geneva,
  1964\relax
\mciteBstWouldAddEndPuncttrue
\mciteSetBstMidEndSepPunct{\mcitedefaultmidpunct}
{\mcitedefaultendpunct}{\mcitedefaultseppunct}\relax
\EndOfBibitem
\bibitem{PDG2022}
Particle Data Group, R.~L. Workman {\em et~al.},
  \ifthenelse{\boolean{articletitles}}{\emph{{\href{http://pdg.lbl.gov/}{Review
  of particle physics}}}, }{}\href{https://doi.org/10.1093/ptep/ptac097}{Prog.\
  Theor.\ Exp.\ Phys.\  \textbf{2022} (2022) 083C01}\relax
\mciteBstWouldAddEndPuncttrue
\mciteSetBstMidEndSepPunct{\mcitedefaultmidpunct}
{\mcitedefaultendpunct}{\mcitedefaultseppunct}\relax
\EndOfBibitem
\bibitem{RevModPhys.90.015003}
S.~L. Olsen, T.~Skwarnicki, and D.~Zieminska,
  \ifthenelse{\boolean{articletitles}}{\emph{Nonstandard heavy mesons and
  baryons: Experimental evidence},
  }{}\href{https://doi.org/10.1103/RevModPhys.90.015003}{Rev.\ Mod.\ Phys.\
  \textbf{90} (2018) 015003}\relax
\mciteBstWouldAddEndPuncttrue
\mciteSetBstMidEndSepPunct{\mcitedefaultmidpunct}
{\mcitedefaultendpunct}{\mcitedefaultseppunct}\relax
\EndOfBibitem
\bibitem{RevModPhys.65.1199}
M.~Anselmino {\em et~al.},
  \ifthenelse{\boolean{articletitles}}{\emph{Diquarks},
  }{}\href{https://doi.org/10.1103/RevModPhys.65.1199}{Rev.\ Mod.\ Phys.\
  \textbf{65} (1993) 1199}\relax
\mciteBstWouldAddEndPuncttrue
\mciteSetBstMidEndSepPunct{\mcitedefaultmidpunct}
{\mcitedefaultendpunct}{\mcitedefaultseppunct}\relax
\EndOfBibitem
\bibitem{JAFFE_2005}
R.~L. Jaffe, \ifthenelse{\boolean{articletitles}}{\emph{Exotica},
  }{}\href{https://doi.org/https://doi.org/10.1016/j.physrep.2004.11.005}{Physics
  Reports \textbf{409} (2005) 1}\relax
\mciteBstWouldAddEndPuncttrue
\mciteSetBstMidEndSepPunct{\mcitedefaultmidpunct}
{\mcitedefaultendpunct}{\mcitedefaultseppunct}\relax
\EndOfBibitem
\bibitem{GronauPhysRevD.93.034020}
M.~Gronau and J.~L. Rosner,
  \ifthenelse{\boolean{articletitles}}{\emph{{$S$-wave nonleptonic hyperon
  decays and
  ${\mathrm{\ensuremath{\Xi}}}_{b}^{\ensuremath{-}}\ensuremath{\rightarrow}{\ensuremath{\pi}}^{\ensuremath{-}}{\mathrm{\ensuremath{\Lambda}}}_{b}$}},
  }{}\href{https://doi.org/10.1103/PhysRevD.93.034020}{Phys.\ Rev.\
  \textbf{D93} (2016) 034020}\relax
\mciteBstWouldAddEndPuncttrue
\mciteSetBstMidEndSepPunct{\mcitedefaultmidpunct}
{\mcitedefaultendpunct}{\mcitedefaultseppunct}\relax
\EndOfBibitem
\bibitem{PenyYuNiuPLB826}
P.-Y. Niu, Q.~Wang, and Q.~Zhao,
  \ifthenelse{\boolean{articletitles}}{\emph{Study of heavy quark conserving
  weak decays in the quark model},
  }{}\href{https://doi.org/https://doi.org/10.1016/j.physletb.2022.136916}{Phys.\
  Lett.\  \textbf{B826} (2022) 136916}\relax
\mciteBstWouldAddEndPuncttrue
\mciteSetBstMidEndSepPunct{\mcitedefaultmidpunct}
{\mcitedefaultendpunct}{\mcitedefaultseppunct}\relax
\EndOfBibitem
\bibitem{Cheng2016}
H.-Y. Cheng {\em et~al.},
  \ifthenelse{\boolean{articletitles}}{\emph{{Heavy-flavor-conserving hadronic
  weak decays of heavy baryons}},
  }{}\href{https://doi.org/10.1007/JHEP03(2016)028}{JHEP \textbf{03} (2016)
  028}, \href{http://arxiv.org/abs/1512.01276}{{\normalfont\ttfamily
  arXiv:1512.01276}}\relax
\mciteBstWouldAddEndPuncttrue
\mciteSetBstMidEndSepPunct{\mcitedefaultmidpunct}
{\mcitedefaultendpunct}{\mcitedefaultseppunct}\relax
\EndOfBibitem
\bibitem{FALLER2015653}
S.~Faller and T.~Mannel, \ifthenelse{\boolean{articletitles}}{\emph{Light-quark
  decays in heavy hadrons},
  }{}\href{https://doi.org/https://doi.org/10.1016/j.physletb.2015.09.072}{Phys.\
  Lett.\  \textbf{B750} (2015) 653}\relax
\mciteBstWouldAddEndPuncttrue
\mciteSetBstMidEndSepPunct{\mcitedefaultmidpunct}
{\mcitedefaultendpunct}{\mcitedefaultseppunct}\relax
\EndOfBibitem
\bibitem{PhysRevD.90.033016}
X.~Li and M.~B. Voloshin, \ifthenelse{\boolean{articletitles}}{\emph{{Decays
  $\Xi_b \to \Lambda_{b} \pi$ and diquark correlations in hyperons}},
  }{}\href{https://doi.org/10.1103/PhysRevD.90.033016}{Phys.\ Rev.\
  \textbf{D90} (2014) 033016},
  \href{http://arxiv.org/abs/1407.2556}{{\normalfont\ttfamily
  arXiv:1407.2556}}\relax
\mciteBstWouldAddEndPuncttrue
\mciteSetBstMidEndSepPunct{\mcitedefaultmidpunct}
{\mcitedefaultendpunct}{\mcitedefaultseppunct}\relax
\EndOfBibitem
\bibitem{Voloshin:2000et}
M.~B. Voloshin, \ifthenelse{\boolean{articletitles}}{\emph{{Weak decays
  $\Xi_Q\to\Lambda_Q\pi$}},
  }{}\href{https://doi.org/10.1016/S0370-2693(00)00150-7}{Phys.\ Lett.\
  \textbf{B476} (2000) 297},
  \href{http://arxiv.org/abs/hep-ph/0001057}{{\normalfont\ttfamily
  arXiv:hep-ph/0001057}}\relax
\mciteBstWouldAddEndPuncttrue
\mciteSetBstMidEndSepPunct{\mcitedefaultmidpunct}
{\mcitedefaultendpunct}{\mcitedefaultseppunct}\relax
\EndOfBibitem
\bibitem{PhysRevD.105.094011}
H.-Y. Cheng and F.~Xu,
  \ifthenelse{\boolean{articletitles}}{\emph{Heavy-flavor-conserving hadronic
  weak decays of charmed and bottom baryons},
  }{}\href{https://doi.org/10.1103/PhysRevD.105.094011}{Phys.\ Rev.\
  \textbf{D105} (2022) 094011}\relax
\mciteBstWouldAddEndPuncttrue
\mciteSetBstMidEndSepPunct{\mcitedefaultmidpunct}
{\mcitedefaultendpunct}{\mcitedefaultseppunct}\relax
\EndOfBibitem
\bibitem{PhysRevD.106.093005}
H.-Y. Cheng, C.-W. Liu, and F.~Xu,
  \ifthenelse{\boolean{articletitles}}{\emph{Heavy-flavor-conserving hadronic
  weak decays of charmed and bottom baryons: An update},
  }{}\href{https://doi.org/10.1103/PhysRevD.106.093005}{Phys.\ Rev.\
  \textbf{D106} (2022) 093005}\relax
\mciteBstWouldAddEndPuncttrue
\mciteSetBstMidEndSepPunct{\mcitedefaultmidpunct}
{\mcitedefaultendpunct}{\mcitedefaultseppunct}\relax
\EndOfBibitem
\bibitem{Shifman:2005wa}
M.~Shifman and A.~Vainshtein,
  \ifthenelse{\boolean{articletitles}}{\emph{{Remarks on diquarks, strong
  binding and a large hidden QCD scale}},
  }{}\href{https://doi.org/10.1103/PhysRevD.71.074010}{Phys.\ Rev.\
  \textbf{D71} (2005) 074010},
  \href{http://arxiv.org/abs/hep-ph/0501200}{{\normalfont\ttfamily
  arXiv:hep-ph/0501200}}\relax
\mciteBstWouldAddEndPuncttrue
\mciteSetBstMidEndSepPunct{\mcitedefaultmidpunct}
{\mcitedefaultendpunct}{\mcitedefaultseppunct}\relax
\EndOfBibitem
\bibitem{Dosch:1988hu}
H.~G. Dosch, M.~Jamin, and B.~Stech,
  \ifthenelse{\boolean{articletitles}}{\emph{{Diquarks, {QCD} sum rules and
  weak decays}}, }{}\href{https://doi.org/10.1007/BF01565139}{Z.\ Phys.\
  \textbf{C42} (1989) 167}\relax
\mciteBstWouldAddEndPuncttrue
\mciteSetBstMidEndSepPunct{\mcitedefaultmidpunct}
{\mcitedefaultendpunct}{\mcitedefaultseppunct}\relax
\EndOfBibitem
\bibitem{Lenz:2014jha}
A.~Lenz, \ifthenelse{\boolean{articletitles}}{\emph{{Lifetimes and heavy quark
  expansion}}, }{}\href{https://doi.org/10.1142/S0217751X15430058}{Int.\ J.\
  Mod.\ Phys.\  \textbf{A30} (2015) 1543005},
  \href{http://arxiv.org/abs/1405.3601}{{\normalfont\ttfamily
  arXiv:1405.3601}}\relax
\mciteBstWouldAddEndPuncttrue
\mciteSetBstMidEndSepPunct{\mcitedefaultmidpunct}
{\mcitedefaultendpunct}{\mcitedefaultseppunct}\relax
\EndOfBibitem
\bibitem{LHCb-PAPER-2015-047}
LHCb collaboration, R.~Aaij {\em et~al.},
  \ifthenelse{\boolean{articletitles}}{\emph{{Evidence for the
  strangeness-changing weak decay \mbox{\decay{\Xibm}{\Lb\pim}}}},
  }{}\href{https://doi.org/10.1103/PhysRevLett.115.241801}{Phys.\ Rev.\ Lett.\
  \textbf{115} (2015) 241801},
  \href{http://arxiv.org/abs/1510.03829}{{\normalfont\ttfamily
  arXiv:1510.03829}}\relax
\mciteBstWouldAddEndPuncttrue
\mciteSetBstMidEndSepPunct{\mcitedefaultmidpunct}
{\mcitedefaultendpunct}{\mcitedefaultseppunct}\relax
\EndOfBibitem
\bibitem{LHCb-PAPER-2018-047}
LHCb collaboration, R.~Aaij {\em et~al.},
  \ifthenelse{\boolean{articletitles}}{\emph{{Measurement of the mass and
  production rate of \Xibm baryons}},
  }{}\href{https://doi.org/10.1103/PhysRevD.99.052006}{Phys.\ Rev.\
  \textbf{D99} (2019) 052006},
  \href{http://arxiv.org/abs/1901.07075}{{\normalfont\ttfamily
  arXiv:1901.07075}}\relax
\mciteBstWouldAddEndPuncttrue
\mciteSetBstMidEndSepPunct{\mcitedefaultmidpunct}
{\mcitedefaultendpunct}{\mcitedefaultseppunct}\relax
\EndOfBibitem
\bibitem{LHCb-PAPER-2020-032}
LHCb collaboration, R.~Aaij {\em et~al.},
  \ifthenelse{\boolean{articletitles}}{\emph{{Observation of a new $\Xibz$
  state}}, }{}\href{https://doi.org/10.1103/PhysRevD.103.012004}{Phys.\ Rev.\
  \textbf{D103} (2021) 012004},
  \href{http://arxiv.org/abs/2010.14485}{{\normalfont\ttfamily
  arXiv:2010.14485}}\relax
\mciteBstWouldAddEndPuncttrue
\mciteSetBstMidEndSepPunct{\mcitedefaultmidpunct}
{\mcitedefaultendpunct}{\mcitedefaultseppunct}\relax
\EndOfBibitem
\bibitem{LHCb-DP-2008-001}
LHCb collaboration, A.~A. Alves~Jr.\ {\em et~al.},
  \ifthenelse{\boolean{articletitles}}{\emph{{The \lhcb detector at the LHC}},
  }{}\href{https://doi.org/10.1088/1748-0221/3/08/S08005}{JINST \textbf{3}
  (2008) S08005}\relax
\mciteBstWouldAddEndPuncttrue
\mciteSetBstMidEndSepPunct{\mcitedefaultmidpunct}
{\mcitedefaultendpunct}{\mcitedefaultseppunct}\relax
\EndOfBibitem
\bibitem{LHCb-DP-2014-002}
LHCb collaboration, R.~Aaij {\em et~al.},
  \ifthenelse{\boolean{articletitles}}{\emph{{LHCb detector performance}},
  }{}\href{https://doi.org/10.1142/S0217751X15300227}{Int.\ J.\ Mod.\ Phys.\
  \textbf{A30} (2015) 1530022},
  \href{http://arxiv.org/abs/1412.6352}{{\normalfont\ttfamily
  arXiv:1412.6352}}\relax
\mciteBstWouldAddEndPuncttrue
\mciteSetBstMidEndSepPunct{\mcitedefaultmidpunct}
{\mcitedefaultendpunct}{\mcitedefaultseppunct}\relax
\EndOfBibitem
\bibitem{LHCb-DP-2014-001}
R.~Aaij {\em et~al.}, \ifthenelse{\boolean{articletitles}}{\emph{{Performance
  of the LHCb Vertex Locator}},
  }{}\href{https://doi.org/10.1088/1748-0221/9/09/P09007}{JINST \textbf{9}
  (2014) P09007}, \href{http://arxiv.org/abs/1405.7808}{{\normalfont\ttfamily
  arXiv:1405.7808}}\relax
\mciteBstWouldAddEndPuncttrue
\mciteSetBstMidEndSepPunct{\mcitedefaultmidpunct}
{\mcitedefaultendpunct}{\mcitedefaultseppunct}\relax
\EndOfBibitem
\bibitem{LHCb-DP-2017-001}
P.~d'Argent {\em et~al.}, \ifthenelse{\boolean{articletitles}}{\emph{{Improved
  performance of the LHCb Outer Tracker in LHC Run 2}},
  }{}\href{https://doi.org/10.1088/1748-0221/12/11/P11016}{JINST \textbf{12}
  (2017) P11016}, \href{http://arxiv.org/abs/1708.00819}{{\normalfont\ttfamily
  arXiv:1708.00819}}\relax
\mciteBstWouldAddEndPuncttrue
\mciteSetBstMidEndSepPunct{\mcitedefaultmidpunct}
{\mcitedefaultendpunct}{\mcitedefaultseppunct}\relax
\EndOfBibitem
\bibitem{LHCb-DP-2012-003}
M.~Adinolfi {\em et~al.},
  \ifthenelse{\boolean{articletitles}}{\emph{{Performance of the \lhcb RICH
  detector at the LHC}},
  }{}\href{https://doi.org/10.1140/epjc/s10052-013-2431-9}{Eur.\ Phys.\ J.\
  \textbf{C73} (2013) 2431},
  \href{http://arxiv.org/abs/1211.6759}{{\normalfont\ttfamily
  arXiv:1211.6759}}\relax
\mciteBstWouldAddEndPuncttrue
\mciteSetBstMidEndSepPunct{\mcitedefaultmidpunct}
{\mcitedefaultendpunct}{\mcitedefaultseppunct}\relax
\EndOfBibitem
\bibitem{LHCb-DP-2012-002}
A.~A. Alves~Jr.\ {\em et~al.},
  \ifthenelse{\boolean{articletitles}}{\emph{{Performance of the LHCb muon
  system}}, }{}\href{https://doi.org/10.1088/1748-0221/8/02/P02022}{JINST
  \textbf{8} (2013) P02022},
  \href{http://arxiv.org/abs/1211.1346}{{\normalfont\ttfamily
  arXiv:1211.1346}}\relax
\mciteBstWouldAddEndPuncttrue
\mciteSetBstMidEndSepPunct{\mcitedefaultmidpunct}
{\mcitedefaultendpunct}{\mcitedefaultseppunct}\relax
\EndOfBibitem
\bibitem{LHCb-DP-2019-001}
R.~Aaij {\em et~al.}, \ifthenelse{\boolean{articletitles}}{\emph{{Performance
  of the LHCb trigger and full real-time reconstruction in Run 2 of the LHC}},
  }{}\href{https://doi.org/10.1088/1748-0221/14/04/P04013}{JINST \textbf{14}
  (2019) P04013}, \href{http://arxiv.org/abs/1812.10790}{{\normalfont\ttfamily
  arXiv:1812.10790}}\relax
\mciteBstWouldAddEndPuncttrue
\mciteSetBstMidEndSepPunct{\mcitedefaultmidpunct}
{\mcitedefaultendpunct}{\mcitedefaultseppunct}\relax
\EndOfBibitem
\bibitem{BBDT}
V.~V. Gligorov and M.~Williams,
  \ifthenelse{\boolean{articletitles}}{\emph{{Efficient, reliable and fast
  high-level triggering using a bonsai boosted decision tree}},
  }{}\href{https://doi.org/10.1088/1748-0221/8/02/P02013}{JINST \textbf{8}
  (2013) P02013}, \href{http://arxiv.org/abs/1210.6861}{{\normalfont\ttfamily
  arXiv:1210.6861}}\relax
\mciteBstWouldAddEndPuncttrue
\mciteSetBstMidEndSepPunct{\mcitedefaultmidpunct}
{\mcitedefaultendpunct}{\mcitedefaultseppunct}\relax
\EndOfBibitem
\bibitem{LHCb-PROC-2015-018}
T.~Likhomanenko {\em et~al.}, \ifthenelse{\boolean{articletitles}}{\emph{{LHCb
  topological trigger reoptimization}},
  }{}\href{https://doi.org/10.1088/1742-6596/664/8/082025}{J.\ Phys.\ Conf.\
  Ser.\  \textbf{664} (2015) 082025}\relax
\mciteBstWouldAddEndPuncttrue
\mciteSetBstMidEndSepPunct{\mcitedefaultmidpunct}
{\mcitedefaultendpunct}{\mcitedefaultseppunct}\relax
\EndOfBibitem
\bibitem{Sjostrand:2007gs}
T.~Sj\"{o}strand, S.~Mrenna, and P.~Skands,
  \ifthenelse{\boolean{articletitles}}{\emph{{A brief introduction to PYTHIA
  8.1}}, }{}\href{https://doi.org/10.1016/j.cpc.2008.01.036}{Comput.\ Phys.\
  Commun.\  \textbf{178} (2008) 852},
  \href{http://arxiv.org/abs/0710.3820}{{\normalfont\ttfamily
  arXiv:0710.3820}}\relax
\mciteBstWouldAddEndPuncttrue
\mciteSetBstMidEndSepPunct{\mcitedefaultmidpunct}
{\mcitedefaultendpunct}{\mcitedefaultseppunct}\relax
\EndOfBibitem
\bibitem{Sjostrand:2006za}
T.~Sj\"{o}strand, S.~Mrenna, and P.~Skands,
  \ifthenelse{\boolean{articletitles}}{\emph{{PYTHIA 6.4 physics and manual}},
  }{}\href{https://doi.org/10.1088/1126-6708/2006/05/026}{JHEP \textbf{05}
  (2006) 026}, \href{http://arxiv.org/abs/hep-ph/0603175}{{\normalfont\ttfamily
  arXiv:hep-ph/0603175}}\relax
\mciteBstWouldAddEndPuncttrue
\mciteSetBstMidEndSepPunct{\mcitedefaultmidpunct}
{\mcitedefaultendpunct}{\mcitedefaultseppunct}\relax
\EndOfBibitem
\bibitem{LHCb-PROC-2010-056}
I.~Belyaev {\em et~al.}, \ifthenelse{\boolean{articletitles}}{\emph{{Handling
  of the generation of primary events in Gauss, the LHCb simulation
  framework}}, }{}\href{https://doi.org/10.1088/1742-6596/331/3/032047}{J.\
  Phys.\ Conf.\ Ser.\  \textbf{331} (2011) 032047}\relax
\mciteBstWouldAddEndPuncttrue
\mciteSetBstMidEndSepPunct{\mcitedefaultmidpunct}
{\mcitedefaultendpunct}{\mcitedefaultseppunct}\relax
\EndOfBibitem
\bibitem{Lange:2001uf}
D.~J. Lange, \ifthenelse{\boolean{articletitles}}{\emph{{The EvtGen particle
  decay simulation package}},
  }{}\href{https://doi.org/10.1016/S0168-9002(01)00089-4}{Nucl.\ Instrum.\
  Meth.\  \textbf{A462} (2001) 152}\relax
\mciteBstWouldAddEndPuncttrue
\mciteSetBstMidEndSepPunct{\mcitedefaultmidpunct}
{\mcitedefaultendpunct}{\mcitedefaultseppunct}\relax
\EndOfBibitem
\bibitem{davidson2015photos}
N.~Davidson, T.~Przedzinski, and Z.~Was,
  \ifthenelse{\boolean{articletitles}}{\emph{{PHOTOS interface in C++:
  Technical and physics documentation}},
  }{}\href{https://doi.org/https://doi.org/10.1016/j.cpc.2015.09.013}{Comp.\
  Phys.\ Comm.\  \textbf{199} (2016) 86},
  \href{http://arxiv.org/abs/1011.0937}{{\normalfont\ttfamily
  arXiv:1011.0937}}\relax
\mciteBstWouldAddEndPuncttrue
\mciteSetBstMidEndSepPunct{\mcitedefaultmidpunct}
{\mcitedefaultendpunct}{\mcitedefaultseppunct}\relax
\EndOfBibitem
\bibitem{Allison:2006ve}
Geant4 collaboration, J.~Allison {\em et~al.},
  \ifthenelse{\boolean{articletitles}}{\emph{{Geant4 developments and
  applications}}, }{}\href{https://doi.org/10.1109/TNS.2006.869826}{IEEE
  Trans.\ Nucl.\ Sci.\  \textbf{53} (2006) 270}\relax
\mciteBstWouldAddEndPuncttrue
\mciteSetBstMidEndSepPunct{\mcitedefaultmidpunct}
{\mcitedefaultendpunct}{\mcitedefaultseppunct}\relax
\EndOfBibitem
\bibitem{Agostinelli:2002hh}
Geant4 collaboration, S.~Agostinelli {\em et~al.},
  \ifthenelse{\boolean{articletitles}}{\emph{{Geant4: A simulation toolkit}},
  }{}\href{https://doi.org/10.1016/S0168-9002(03)01368-8}{Nucl.\ Instrum.\
  Meth.\  \textbf{A506} (2003) 250}\relax
\mciteBstWouldAddEndPuncttrue
\mciteSetBstMidEndSepPunct{\mcitedefaultmidpunct}
{\mcitedefaultendpunct}{\mcitedefaultseppunct}\relax
\EndOfBibitem
\bibitem{LHCb-PROC-2011-006}
M.~Clemencic {\em et~al.}, \ifthenelse{\boolean{articletitles}}{\emph{{The
  \lhcb simulation application, Gauss: Design, evolution and experience}},
  }{}\href{https://doi.org/10.1088/1742-6596/331/3/032023}{J.\ Phys.\ Conf.\
  Ser.\  \textbf{331} (2011) 032023}\relax
\mciteBstWouldAddEndPuncttrue
\mciteSetBstMidEndSepPunct{\mcitedefaultmidpunct}
{\mcitedefaultendpunct}{\mcitedefaultseppunct}\relax
\EndOfBibitem
\bibitem{LHCb-DP-2018-004}
D.~M{\"u}ller, M.~Clemencic, G.~Corti, and M.~Gersabeck,
  \ifthenelse{\boolean{articletitles}}{\emph{{ReDecay: A novel approach to
  speed up the simulation at LHCb}},
  }{}\href{https://doi.org/10.1140/epjc/s10052-018-6469-6}{Eur.\ Phys.\ J.\
  \textbf{C78} (2018) 1009},
  \href{http://arxiv.org/abs/1810.10362}{{\normalfont\ttfamily
  arXiv:1810.10362}}\relax
\mciteBstWouldAddEndPuncttrue
\mciteSetBstMidEndSepPunct{\mcitedefaultmidpunct}
{\mcitedefaultendpunct}{\mcitedefaultseppunct}\relax
\EndOfBibitem
\bibitem{Breiman}
L.~Breiman, J.~H. Friedman, R.~A. Olshen, and C.~J. Stone, {\em Classification
  and regression trees}, Wadsworth international group, Belmont, California,
  USA, 1984\relax
\mciteBstWouldAddEndPuncttrue
\mciteSetBstMidEndSepPunct{\mcitedefaultmidpunct}
{\mcitedefaultendpunct}{\mcitedefaultseppunct}\relax
\EndOfBibitem
\bibitem{AdaBoost}
Y.~Freund and R.~E. Schapire, \ifthenelse{\boolean{articletitles}}{\emph{A
  decision-theoretic generalization of on-line learning and an application to
  boosting}, }{}\href{https://doi.org/10.1006/jcss.1997.1504}{J.\ Comput.\
  Syst.\ Sci.\  \textbf{55} (1997) 119}\relax
\mciteBstWouldAddEndPuncttrue
\mciteSetBstMidEndSepPunct{\mcitedefaultmidpunct}
{\mcitedefaultendpunct}{\mcitedefaultseppunct}\relax
\EndOfBibitem
\bibitem{TMVA4}
A.~Hoecker {\em et~al.}, \ifthenelse{\boolean{articletitles}}{\emph{{TMVA 4 ---
  Toolkit for Multivariate Data Analysis with ROOT. Users Guide.}},
  }{}\href{http://arxiv.org/abs/physics/0703039}{{\normalfont\ttfamily
  arXiv:physics/0703039}}\relax
\mciteBstWouldAddEndPuncttrue
\mciteSetBstMidEndSepPunct{\mcitedefaultmidpunct}
{\mcitedefaultendpunct}{\mcitedefaultseppunct}\relax
\EndOfBibitem
\bibitem{Pivk:2004ty}
M.~Pivk and F.~R. Le~Diberder,
  \ifthenelse{\boolean{articletitles}}{\emph{{sPlot: A statistical tool to
  unfold data distributions}},
  }{}\href{https://doi.org/10.1016/j.nima.2005.08.106}{Nucl.\ Instrum.\ Meth.\
  \textbf{A555} (2005) 356},
  \href{http://arxiv.org/abs/physics/0402083}{{\normalfont\ttfamily
  arXiv:physics/0402083}}\relax
\mciteBstWouldAddEndPuncttrue
\mciteSetBstMidEndSepPunct{\mcitedefaultmidpunct}
{\mcitedefaultendpunct}{\mcitedefaultseppunct}\relax
\EndOfBibitem
\bibitem{LHCb-DP-2018-001}
R.~Aaij {\em et~al.}, \ifthenelse{\boolean{articletitles}}{\emph{{Selection and
  processing of calibration samples to measure the particle identification
  performance of the LHCb experiment in Run 2}},
  }{}\href{https://doi.org/10.1140/epjti/s40485-019-0050-z}{Eur.\ Phys.\ J.\
  Tech.\ Instr.\  \textbf{6} (2019) 1},
  \href{http://arxiv.org/abs/1803.00824}{{\normalfont\ttfamily
  arXiv:1803.00824}}\relax
\mciteBstWouldAddEndPuncttrue
\mciteSetBstMidEndSepPunct{\mcitedefaultmidpunct}
{\mcitedefaultendpunct}{\mcitedefaultseppunct}\relax
\EndOfBibitem
\bibitem{Skwarnicki:1986xj}
T.~Skwarnicki, {\em {A study of the radiative cascade transitions between the
  Upsilon-prime and Upsilon resonances}}, PhD thesis, Institute of Nuclear
  Physics, Krakow, 1986,
  {\href{http://inspirehep.net/record/230779/}{DESY-F31-86-02}}\relax
\mciteBstWouldAddEndPuncttrue
\mciteSetBstMidEndSepPunct{\mcitedefaultmidpunct}
{\mcitedefaultendpunct}{\mcitedefaultseppunct}\relax
\EndOfBibitem
\bibitem{LHCb-PAPER-2014-021}
LHCb collaboration, R.~Aaij {\em et~al.},
  \ifthenelse{\boolean{articletitles}}{\emph{{Precision measurement of the mass
  and lifetime of the \Xibz baryon}},
  }{}\href{https://doi.org/10.1103/PhysRevLett.113.032001}{Phys.\ Rev.\ Lett.\
  \textbf{113} (2014) 032001},
  \href{http://arxiv.org/abs/1405.7223}{{\normalfont\ttfamily
  arXiv:1405.7223}}\relax
\mciteBstWouldAddEndPuncttrue
\mciteSetBstMidEndSepPunct{\mcitedefaultmidpunct}
{\mcitedefaultendpunct}{\mcitedefaultseppunct}\relax
\EndOfBibitem
\bibitem{Wilks:1938dza}
S.~S. Wilks, \ifthenelse{\boolean{articletitles}}{\emph{{The large-sample
  distribution of the likelihood ratio for testing composite hypotheses}},
  }{}\href{https://doi.org/10.1214/aoms/1177732360}{Ann.\ Math.\ Stat.\
  \textbf{9} (1938) 60}\relax
\mciteBstWouldAddEndPuncttrue
\mciteSetBstMidEndSepPunct{\mcitedefaultmidpunct}
{\mcitedefaultendpunct}{\mcitedefaultseppunct}\relax
\EndOfBibitem
\bibitem{LHCb-DP-2013-002}
LHCb collaboration, R.~Aaij {\em et~al.},
  \ifthenelse{\boolean{articletitles}}{\emph{{Measurement of the track
  reconstruction efficiency at LHCb}},
  }{}\href{https://doi.org/10.1088/1748-0221/10/02/P02007}{JINST \textbf{10}
  (2015) P02007}, \href{http://arxiv.org/abs/1408.1251}{{\normalfont\ttfamily
  arXiv:1408.1251}}\relax
\mciteBstWouldAddEndPuncttrue
\mciteSetBstMidEndSepPunct{\mcitedefaultmidpunct}
{\mcitedefaultendpunct}{\mcitedefaultseppunct}\relax
\EndOfBibitem
\bibitem{LHCb-DP-2012-004}
R.~Aaij {\em et~al.}, \ifthenelse{\boolean{articletitles}}{\emph{{The \lhcb
  trigger and its performance in 2011}},
  }{}\href{https://doi.org/10.1088/1748-0221/8/04/P04022}{JINST \textbf{8}
  (2013) P04022}, \href{http://arxiv.org/abs/1211.3055}{{\normalfont\ttfamily
  arXiv:1211.3055}}\relax
\mciteBstWouldAddEndPuncttrue
\mciteSetBstMidEndSepPunct{\mcitedefaultmidpunct}
{\mcitedefaultendpunct}{\mcitedefaultseppunct}\relax
\EndOfBibitem
\end{mcitethebibliography}

\newpage
\centerline
{\large\bf LHCb collaboration}
\begin
{flushleft}
\small
R.~Aaij$^{33}$\lhcborcid{0000-0003-0533-1952},
A.S.W.~Abdelmotteleb$^{52}$\lhcborcid{0000-0001-7905-0542},
C.~Abellan~Beteta$^{46}$,
F.~Abudin{\'e}n$^{52}$\lhcborcid{0000-0002-6737-3528},
T.~Ackernley$^{56}$\lhcborcid{0000-0002-5951-3498},
B.~Adeva$^{42}$\lhcborcid{0000-0001-9756-3712},
M.~Adinolfi$^{50}$\lhcborcid{0000-0002-1326-1264},
P.~Adlarson$^{78}$\lhcborcid{0000-0001-6280-3851},
H.~Afsharnia$^{10}$,
C.~Agapopoulou$^{44}$\lhcborcid{0000-0002-2368-0147}
% ...

{\footnotesize \it

% ...
$^{10}$Universit{\'e} Clermont Auvergne, CNRS/IN2P3, LPC, Clermont-Ferrand, France\\
% ...
$^{33}$Nikhef National Institute for Subatomic Physics, Amsterdam, Netherlands\\
% ...
$^{42}$Instituto Galego de F{\'\i}sica de Altas Enerx{\'\i}as (IGFAE), Universidade de Santiago de Compostela, Santiago de Compostela, Spain\\
% ...
$^{44}$European Organization for Nuclear Research (CERN), Geneva, Switzerland\\
% ...
$^{46}$Physik-Institut, Universit{\"a}t Z{\"u}rich, Z{\"u}rich, Switzerland\\
% ...
$^{50}$H.H. Wills Physics Laboratory, University of Bristol, Bristol, United Kingdom\\
% ...
$^{52}$Department of Physics, University of Warwick, Coventry, United Kingdom\\
% ...
$^{55}$School of Physics and Astronomy, University of Glasgow, Glasgow, United Kingdom\\
$^{56}$Oliver Lodge Laboratory, University of Liverpool, Liverpool, United Kingdom\\
% ...
$^{78}$Department of Physics and Astronomy, Uppsala University, Uppsala, Sweden, associated to $^{55}$\\
% ...
}
\end{flushleft}

\end{document}